\documentclass[format=acmsmall, review=false, screen=true]{acmart}
\usepackage{dcolumn}
\usepackage{url}
\usepackage{booktabs} %
\usepackage{graphics}   
\usepackage{comment}
\usepackage{subcaption}
\usepackage{caption}
\usepackage{fancychapters} 
\usepackage{xspace}
\usepackage{amsmath}
\usepackage[final]{changes}

\usepackage[ruled]{algorithm2e} %

\newcommand{\communityname}[1]{{\sf #1}\xspace}
\newcommand{\topicname}[1]{{\it ``#1''}\xspace}
\newcommand{\variablename}[1]{\text{\it\small #1}}
\newcommand{\para}[1]{\noindent{\bf #1}\xspace}
\newcommand{\secref}[1]{Section \ref{#1}\xspace}

\SetAlFnt{\small}
\SetAlCapFnt{\small}
\SetAlCapNameFnt{\small}
\SetAlCapHSkip{0pt}
\IncMargin{-\parindent}

\newcolumntype{L}{D{.}{.}{2,5}}

\setcopyright{acmcopyright}
\acmJournal{PACMHCI}
\acmYear{2018} 
\acmVolume{2} 
\acmNumber{CSCW} 
\acmArticle{197} 
\acmMonth{1} 
\acmPrice{15.00}
\acmDOI{10.1145/3274466}

\received{April 2018}
\received[revised]{July 2018}
\received[accepted]{August 2018}

\graphicspath{ {graphics/} }

\begin{document}
\title[Characterizing Online Fan Communities of the NBA Teams]{``This is why we play'': Characterizing Online Fan Communities of the NBA Teams}

\author{Jason Shuo Zhang, Chenhao Tan, and Qin Lv}
\affiliation{%
  \institution{University of Colorado Boulder}
  \streetaddress{1111 Engineering Dr}
  \city{Boulder}
  \state{CO}
  \postcode{80309}
  \country{USA}}

\thanks{We thank anonymous reviewers for helpful comments and discussions. 
We thank Jason Baumgartner and Jack Hessel for sharing the dataset that enabled this research.
We thank Scott Holman from the CU Boulder Writing Center for the very detailed proofreading.
This work is supported in part by the US National Science
Foundation (NSF) through grant CNS 1528138.}

\begin{abstract}
Professional sports constitute an important part of people's modern life.
People spend substantial amounts of time and money supporting their favorite players and teams, 
and sometimes even riot after \deleted{unsuccessful}games.
However, how team performance affects fan behavior remains understudied at a large scale.
As almost every notable professional team has its own online fan community, these communities provide great opportunities for investigating this research question.
In this work, 
we provide the first large-scale characterization of online fan communities of professional sports teams.

Since user behavior in these online fan communities is inherently connected to game events and team performance,
we construct a unique dataset that combines \replaced{1.5M}{1.4M} posts and 43M comments in NBA-related communities on Reddit with statistics that document team performance in the NBA.
We analyze the impact of team performance on fan behavior both at the game level and the season level.
First, we study how team performance in a game relates to user activity during that game.
We find that surprise plays an important role: the fans of the top teams are more active when their teams 
lose and so are the fans of the bottom teams in an 
unexpected win.
Second, we study fan behavior over consecutive seasons and show that strong team performance is associated with fans of low loyalty, likely due to ``bandwagon fans.''
Fans of the bottom teams tend to discuss their team's future such as young talents in the roster, which may help them stay optimistic during adversity.
Our results not only contribute to understanding the interplay between online sports communities and offline context
but also provide significant insights into sports management.
\end{abstract}

\ccsdesc[500]{Applied computing~Law, social and behavioral sciences}

\keywords{online fan communities, team performance, professional sports, NBA}

\maketitle

\renewcommand{\shortauthors}{Author(s) Anonymous for Review}

\section{Introduction}

\textit{Our [the Los Angeles Lakers'] collective success having forged some kind of unity in this huge and normally fragmented metropolis, it cuts across cultural and class lines.}
\\[5pt]
\rightline{{\rm --- Kareem Abdul-Jabbar, an NBA \replaced{Hall of Famer}{legend}.}}\\[3pt]

Professional sports not only involve competitions among athletes, but also attract fans to attend the games, watch broadcasts, and consume related products \cite{wenner1989media}.
For instance, the 2017 final game of the National Basketball Association (NBA) %
attracted 20 million TV viewers \cite{statista:nba}; a 30-second commercial during the Super Bowl cost around 4.5 million dollars in 2015 and these commercials have become an integral part of American culture \cite{wiki:NFL}.\footnote{Most influential Super Bowl commercials: \url{http://time.com/4653281/super-bowl-ads-commercials-most-influential-time/}.}
Fans of sports teams can be very emotionally invested and treat
fans of rival teams almost as enemies, which can even lead to violence \cite{roadburg1980factors}.

Such excitement towards professional sports extends to online communities.
A notable example is \communityname{/r/NBA} ~\added{on Reddit}, which attracts over a million subscribers and has become one of the most active subreddits on Reddit, a popular community-driven website 
\cite{redditlist}. 
\citeauthor{rnba}, a sports writer, has suggested that online fan communities are gradually replacing the need for sports blogs and even larger media outlets altogether \cite{rnba}. 
The growth of online fan communities thus provides exciting opportunities for studying fan behavior in professional sports at a large scale.
It is important to recognize that fan behavior is driven by sports events, including sports games, player transfers between teams, and even a comment from a team manager.
The dynamic nature of sports games indicates that discussions in online fan communities may echo the development in games, analogous to the waves of excitement in a stadium.
Therefore, our goal in this paper is to characterize {\em online} fan communities in the context of {\em offline} games and team performance.

To do that, we build a large-scale dataset of online fan communities from Reddit with \replaced{479K}{477K} users, \replaced{1.5M}{1.4M} posts, and 43M comments, as well as statistics that document offline games and team performance.\footnote{The dataset is available at \url{http://jasondarkblue.com/papers/CSCW2018NBADataset_README.txt}.}
We choose Reddit as a testbed because 1) Reddit has explicit communities for every NBA team, which allows us to compare the differences between winning teams and losing teams; and 
2) Reddit is driven by fan communities, e.g., the ranking of posts is determined by upvotes and downvotes of community members.
In comparison, team officials can have a great impact on a team's official Twitter account and Facebook page.

\para{Organization and highlights.} We first \added{summarize related work (\secref{sec:related}) and then} provide an overview of the NBA fan communities on Reddit as well as necessary background knowledge regarding the NBA games (\secref{sec:data}).
We demonstrate the seasonal patterns in online fan communities and how they align with the NBA season in the offline world.
We further characterize the discussions in these NBA fan communities using topic modeling. 

We investigate three research questions in the rest of the paper.
First, we study the relation between team performance in a game and this game's associated fan activity in online fan communities.
We are able to identify game threads that are posted to facilitate discussions during NBA games.
These game threads allow us to examine the short-term impact of team performance on fan behavior.
We demonstrate intriguing contrasts between top teams and bottom teams: user activity increases when top teams lose and bottom teams win.
Furthermore, close games with small point differences are associated with higher user activity levels.

Second, we examine how team performance influences fan loyalty in online communities beyond a single game.
It is important for professional teams to acquire and maintain a strong fan 
base that provides consistent support and consumes team-related products. 
Understanding fan loyalty is thus a central research question in the literature of sports management \cite{dwyer2011divided,stevens2012influence, yoshida2015predicting,doyle2017there}.
For instance, ``bandwagon fan'' refers to a person who follows the tide and supports teams with recent success.
Top teams may have lower fan loyalty due to the existence of many bandwagon fans.
Our results validate this hypothesis by using user retention to measure fan loyalty.
We also find that a team's fan loyalty is correlated with the team's improvement over a season and with the average age of 
the roster.

Third, we turn to the content in online fan communities to understand the impact of team performance on the topics of discussion.
Prior studies show that a strong fan base can minimize the effect of a team's short-term \added{(poor)} performance on its long-term success
\cite{sutton1997creating,shank2014sports}. 
To foster fan identification in teams with poor performance, 
fans may shift the focus from current failure to future success and ``\textit{trust the process}''\footnote{A mantra that reflects Philadelphia 76ers' identity \cite{trusttheprocess} 76ers went through a streak of losing seasons to get top talents in draft-lottery and rebuild the team.} \cite{doyle2017there,campbell2004beyond,jones2000model}.
Discussions in online fan communities enable quantitative analysis of such \added{a} hypothesis.
\replaced{We show that fans of the top teams are more likely to discuss \topicname{season prospects,}
whiles fans of the bottom teams are more likely to discuss \topicname{future.}
Here \topicname{future} refers to the assets that a team has, including talented young players, 
draft picks, and salary space, 
which can potentially prepare the team for future success in the following seasons.}
{We show that fans of bottom teams are more likely to discuss \topicname{future,} while fans of top teams are more likely to discuss \topicname{season prospects}.}

We \deleted{summarize additional related work in \secref{sec:related} and} offer concluding discussions in \secref{sec:conclusion}.
Our work develops the first step towards studying fan behavior in professional sports using online fan communities\deleted{and presents valuable insights for understanding both online communities and sports management.} \added{and provides implications for online communities and sports management. 
For online communities, our results highlight the importance of understanding online behavior in the offline context.
Such offline context can influence the topics of discussion, the activity patterns, and users' decisions to stay or leave.
For sports management, our work reveals strategies for developing a strong fan base such as shifting the topics of discussion and leveraging unexpected wins and potential future success.}

\section{Related Work}
\label{sec:related}

In this section, we survey prior research mainly in two areas related to the work presented in this paper: 
online communities and sports fan behavior.

\subsection{Online Communities}

The proliferation of online communities has enabled a rich body of research in understanding group formation and community dynamics \cite{Backstrom:2006:GFL:1150402.1150412,Ren:07,Kairam:2012:LDO:2124295.2124374,Kim:2000:CBW:518514}.
Most relevant to our work are studies that investigate how external factors affect user behavior in online communities \cite{palen2016crisis,starbird2010chatter,romero2016social,zhang2017shocking}.
\citet{palen2016crisis} provide an overview of studies on social media behavior in response to natural disasters and point out limits of social media data for addressing emergency management.
\citet{romero2016social} find that communication networks between traders ``turtle'' up during shocks in stock price and reveal relations between social network structure and collective behavior.
Other offline events studies include the dynamics of breaking news \cite{keegan2013hot,keegan2012staying,leavitt2014upvoting}, celebrity death \cite{keegan2015coordination,gach2017control}, and Black Lives Matter \cite{twyman2017black,Stewart:2017:DLC:3171581.3134920}. 
This literature illustrates that online communities do not only exist in the virtual world. 
They are usually deeply embedded in the offline context in our daily life.

Another relevant line of work examines user engagement in multiple communities and in particular, user loyalty 
\cite{tan2015all,zhang2017community,hamilton2017loyalty}.
\citet{hamilton2017loyalty} operationalize loyalty in the context of multi-community engagement and consider users loyal to a community if they consistently prefer the community over all others.
They show that loyal users employ language that signals collective identity and their loyalty can be predicted from their first interactions.

Reddit has attracted significant interest from researchers in the past few years due to its growing importance.
Many aspects and properties of Reddit have been extensively studied, including user and subreddit lifecycle in online
platforms \cite{tan2015all,newell2016user}, hate speech \cite{chandrasekharan2017you,chandrasekharan2017bag,saleemweb},
interaction and conflict between subreddits \cite{kumar2018community,tan:18}, 
and its relationship with other web sources \cite{vincent2018examining,newell2016user}.
Studies have also explored the impacts of certain Reddit evolutions and policy changes 
on user behaviors. 
Notable events include pre-default subreddit \cite{lin2017better} and
Reddit unrest \cite{newell2016user,chandrasekharan2017you,matias2016going}.

Our work examines a special set of online communities that derived from professional sports teams.
As a result, regular sports games and team performance are central for understanding these communities and user loyalty in these communities.
Different from prior studies, we focus on the impact of team performance on user behavior in online fan communities.

\subsection{Sports Fan Behavior}
As it is crucial for a sports team to foster a healthy and strong fan base, extensive studies in sports management have studied fan behavior.
Researchers have studied factors that affect purchasing behavior of sports fans
\cite{smith2007travelling, wann2008motivational, trail2001motivation}, including psychometric properties and fan motivation.
A few studies also build predictive models of fan loyalty \cite{bee2010exploring, yoshida2015predicting}.
\citet{bee2010exploring} suggest that fan attraction, involvement, psychological commitment, and resistance can be predictors of fan behavioral loyalty. 
\added{\citet{dolton12018football} estimate  
that the happiness that fans feel when their team
wins is outweighted by the sadness that strikes 
when their team loses by a factor or two.}
\citet{yoshida2015predicting} build regression models based on attitudinal processes to predict behavioral loyalty. 
The potential influence of mobile technology on sports spectators is also examined 
from different angles \cite{torrez2012look,ludvigsen2010designing,jacucci2005supporting}.
\citet{torrez2012look} describes survey results that suggest the current usage of mobile technology 
among college sports fans. The work by \citet{ludvigsen2010designing} examines the potential 
of interactive technologies for active spectating at sporting events.
Most relevant to our work are studies related to fan identification \cite{campbell2004beyond,doyle2017there,dwyer2011divided,hirt1992costs,hyatt2015using,jones2000model,stevens2012influence,sutton1997creating,wann2002using} and we have discussed them to formulate our hypotheses.
These studies usually employ qualitative methods through interviews or small-scale surveys.

It is worth noting that fan behavior can differ depending on the environment.
\citet{cottingham2012interaction} demonstrates the difference in emotional energy between fans in sports bars and those attending the game in the stadium. 
In our work, we focus on online communities, which are an increasingly important platform for sports fans.
These online fan communities also allow us to study team performance and fan behavior at a much larger scale than all existing studies.

\section{An Overview of NBA Fan Communities on Reddit}
\label{sec:data}

Our main dataset is derived from NBA-related communities on Reddit, a popular website organized by communities where users can submit posts and make comments.
A community on Reddit is referred to as a \textit{subreddit}. We use community and subreddit interchangeably in this paper.
We first introduce the history of NBA-related subreddits and then provide an overview of activity levels and discussions in these subreddits.

\subsection{NBA-related Subreddits}
\begin{table}[]
\centering
\begin{tabular}{lrr}
\toprule
         & \communityname{/r/NBA} & \added{Average of team subreddits (std)} \\
\midrule
\added{\#users}    & 400K     & 13K (8K)         \\ 
\#posts    & 847K     & 24K (16K)          \\ 
\#comments & 33M      & 328K (282K)        \\
\bottomrule
\end{tabular}
\caption{Dataset Statistics. There are in total 30 teams in the NBA league. 
\replaced{\#users refers to the number of unique users who posted/commented in the subreddit.}
{``Team subreddits'' statistics are aggregated across all the NBA team subreddits.}}
\label{tab:stats}
\end{table}

On Reddit, the league subreddit \communityname{/r/NBA} is for NBA fans to discuss 
anything that 
happened in the entire league, ranging from a game to gossip related to a player. 
There are 30 teams in total in the NBA league, and 
each team's subreddit is for fans to discuss team-specific topics. 
Each subreddit has multiple moderators to make sure posts are relevant to the subreddit's theme. 
We collected posts and comments in these 31 subreddits (\communityname{/r/NBA} + 30 NBA team subreddits)
from pushshift.io~\cite{pushshift} 
from the beginning of each subreddit until October 2017.~\footnote{A small amount of data is missing
due to scraping errors and other unknown reasons with this dataset \cite{gaffney2018caveat}.
We checked the sensitivity of our results to missing posts with a dataset provided by J.Hessel;
our results in this paper do not change after accounting for them. } 
The overall descriptive statistics of our dataset are shown in Table~\ref{tab:stats}.

\begin{figure}
    \begin{subfigure}[t]{0.46\textwidth}
        \includegraphics[width=\textwidth]{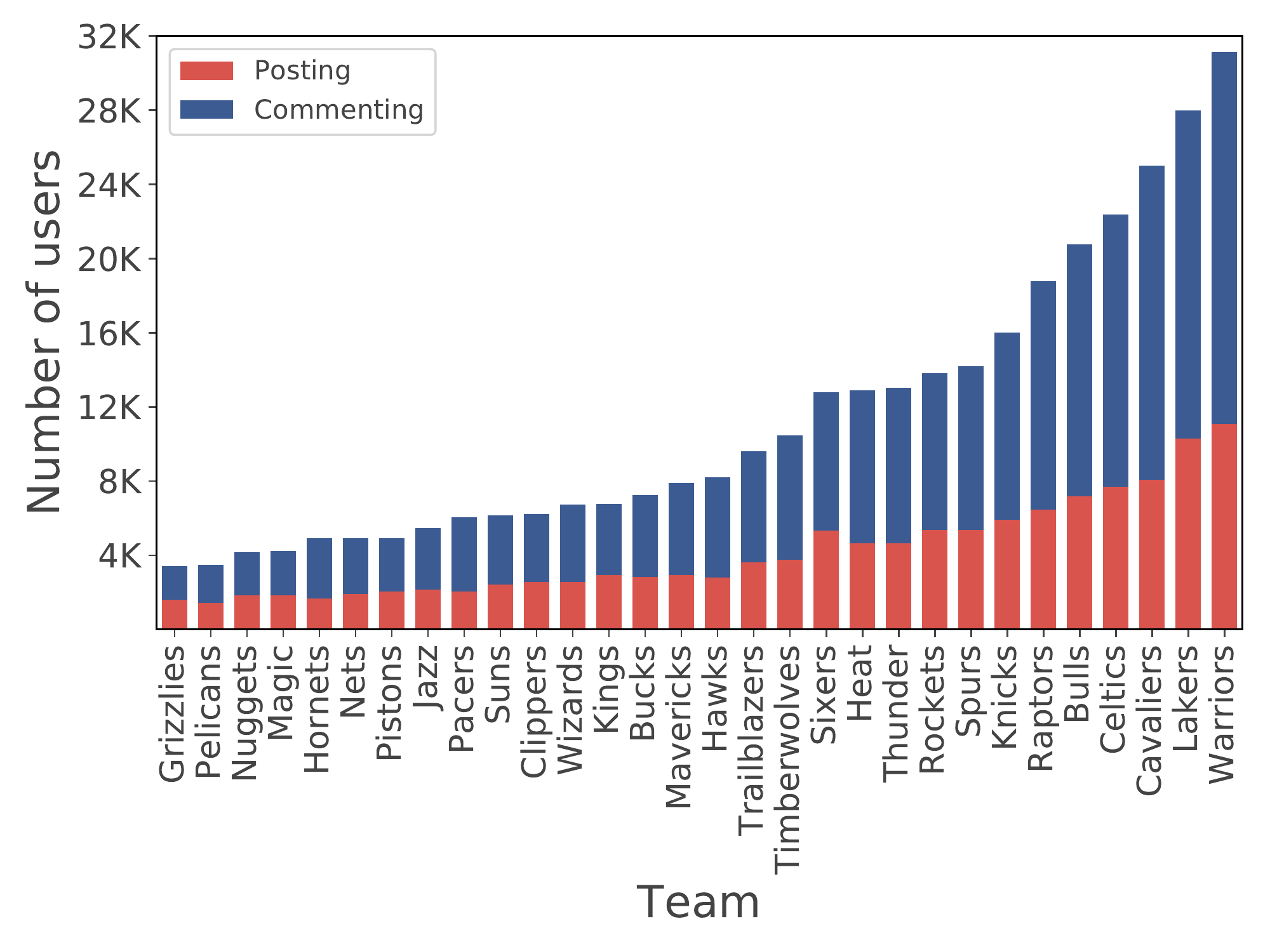}
        \caption{Number of users in team subreddits.}
        \label{fig:user}
    \end{subfigure}
    \hfill
    \begin{subfigure}[t]{0.46\textwidth}
        \includegraphics[width=\textwidth]{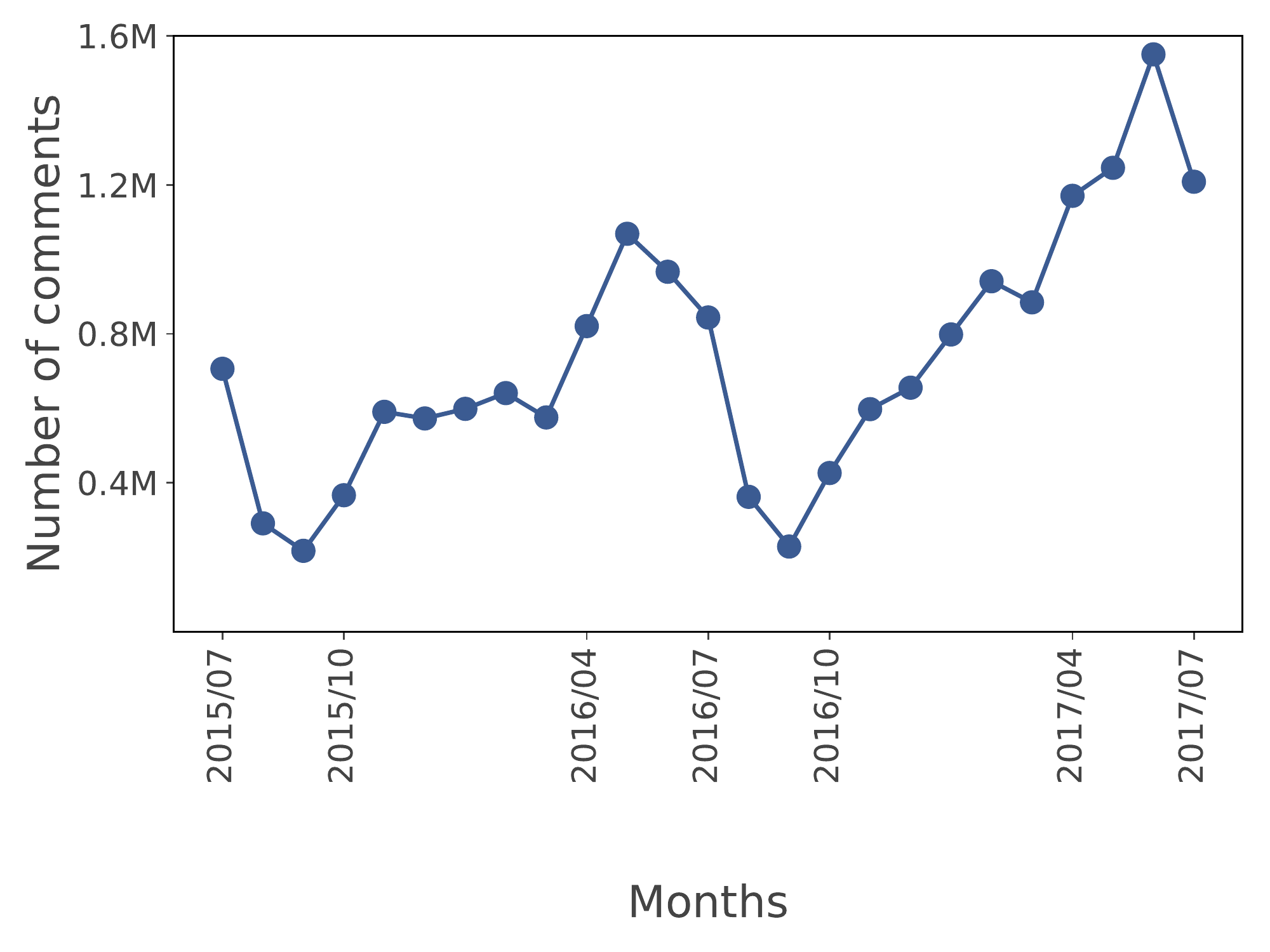}
        \caption{User activity in \communityname{/r/NBA} by month.}
        \label{fig:comment}
    \end{subfigure}
    \caption{Figure~\ref{fig:user} shows the number of users that post and comment in team subreddits. \communityname{/r/Warriors}, \communityname{/r/Lakers}, \communityname{/r/Cavaliers} 
    are the top three subreddits in both posting and commenting. 
    The average number of users across all teams is 4,166 for posting 
    and 11,320 for commenting.
    Figure~\ref{fig:comment} shows user activity level in \communityname{/r/NBA} by month. During the off season (July-mid October), user activity decreases sharply, as no games are played during this period. 
    Then in the regular season (late October to next March),  user activity increases steadily. The activity of \communityname{/r/NBA} 
    peaks in May and June, as the championship games happen in these two months. 
    }
\end{figure}

\noindent\textbf{A brief history of the NBA-related subreddits on Reddit.} 
NBA-related subreddits have thrived since January 2008, when Reddit released a new policy to allow users to create their own subreddit. 
The Lakers' and the Celtics' subreddits were created by fans in 2008, and they are the first two NBA teams to have their team subreddits.
These two teams are also widely acknowledged as the most successful franchises 
in the history of the NBA league~\cite{mostsucessful}. 
It is also worth 
noting
that these two teams' subreddits were created even before \communityname{/r/NBA}, 
which 
was created
at the end of 2008. 
For the remaining 28 teams, 14 of their subreddits were created by users in 2010, and the other 14 were created in 2011. 
Moreover, three teams' subreddit names have changed. 
The Pelicans' subreddit changed their subreddit name from \communityname{/r/Hornets} to \communityname{/r/Nolapelicans} and 
the Hornets' subreddit from \communityname{/r/Charlottebobcats} to \communityname{/r/Charlottehornets} because these two 
teams changed their official team names. Additionally, the Rockets' subreddit shortened its name from \communityname{/r/Houstonrockets} to \communityname{/r/Rockets}. 
To rebuild each team's complete subreddit history, 
we combined posts and comments in these three teams' old and new subreddits. 
Figure~\ref{fig:user} presents the number of users that post and comment in each team subreddit.

\subsection{NBA Season Structure Reflected in Reddit Activity}

As discussed in the introduction, fan behavior in NBA-related subreddits is influenced by offline events.
In particular, the NBA runs in seasons, and seasonal patterns are reflected on Reddit.
To show that, we start with a brief introduction of the NBA.
30 teams in the NBA are divided into two conferences (East and West).
In each season, teams play with each other to compete for the final championship.
\replaced{The following three time segments make up a complete NBA season:}
{A complete season can be divided into three periods:}
\footnote{Please see the NBA's official description for more details \cite{NBARule}.}
\begin{itemize}
\item \textbf{Off season}: from July to mid October. 
There are no games in this
period.\footnote{There are Summer League games and preseason games played in this period, 
but the results don't count in season record.}
Every team is allowed to draft young players, 
sign free agents, and trade players with other teams. 
The bottom teams in the last season get the top positions when drafting young players, which hopefully leads to a long-term balance between teams in the league.
The goal of the off season for each team is to improve its overall competitiveness for the coming season.
\item \textbf{Regular season}: from late October to middle of April. 
Regular season games occur in this period. 
Every team has 82 games scheduled during this time, 
41 home games and 41 away games. 
A team's regular season record is used for playoff qualification and seeding.
\item \textbf{Playoff season}: from the end of regular season to June.
16 teams 
(the top 8 from the Western conference and the top 8 from the Eastern conference) 
play knockouts in each conference 
and compete for the conference championship. 
The champion of the Western Conference and the Eastern Conference 
play the final games to win the final championship.
\end{itemize}
Given the structure, a complete NBA season spans two calendar years. 
In this paper, for simplicity and clarity, we refer to a specific season by the calendar year when it ends. 
For instance, the official 2016-2017 NBA season is referred to as {\em the 2017 season} throughout the paper. 

User activity in NBA-related subreddits is driven by the structure of the NBA season. 
As an example, Figure~\ref{fig:comment} shows user activity in \communityname{/r/NBA} by month in the 2016 and 2017 season. 
From July to September, user activity decreases sharply as 
there are no games during this period. 
Then from October to the next March, the number of comments increases steadily as the regular season unfolds. 
According to the NBA rules, every game in the regular season carries the same weight for playoff qualification, the games in October should be equally important as the games in March.
However, 
fans are much more active \added{on Reddit as it gets} closer to the end of the regular season  because they deem these games ``more critical.''
This may be explained by the ``deadline pressure'' phenomenon in psychology~\cite{ariely2008predictably}. 
\added{This circumstance has also been observed in other sports. 
For instance, \citet{paton2005attendance} illustrate that the attendance of domestic cricket leagues
in England and Wales is much higher in the later segment of the season than the earlier segment.
\citet{hogan2017analysing} find that the possibility of the home team reaching the knock-out stage
had a significant positive impact on attendance in the European Rugby Cup.}
We also find that user activity drops a little bit in April in both the 2016 and 2017 season. 
One possible explanation is that as the regular season is ending,
fans of the bottom teams that clearly cannot make the playoffs 
reduce their activity during this period. 
After that, 
the volume of comments increases dramatically during the playoff games.
The activity of \communityname{/r/NBA} peaks in May and June, when the conference championship and final championship games happen.

\subsection{Topic analysis}

To understand what fans are generally talking about in NBA-related subreddits, 
we use Latent Dirichlet Allocation (LDA)~\cite{blei2003latent}, 
a widely used topic modeling method, to analyze user comments. 
We treat each comment as a document and use all the comments in \communityname{/r/NBA} to train a LDA model
with the Stanford Topic Modeling Toolbox~\cite{stanfordlda}. 
We choose the number of topics based on perplexity scores~\cite{wallach2009evaluation}.
The perplexity score drops significantly \added{when the topic number increases} from 5 to 15, 
but 
does not change much from 15 to 50, all within 1370-1380 range.
Therefore, we use 15 topics in this paper. 
Table~\ref{tab:topic} shows the top five topics with the greatest average topic weight and the
top ten weighted words in each topic.
\added{Two authors, who are NBA fans and active users on \communityname{/r/NBA}, manually assigned a label to each of the five most frequent topics based on the top words in each topic.}
\replaced{Each label}{We name each topic} in Table~\ref{tab:topic} 
\replaced{summarizes}{to summarize} the topic's gist, and 
the five \replaced{labels}{names} are \topicname{personal opinion,} 
\topicname{game strategy,} \topicname{season prospects,} \topicname{future,} and \topicname{game stats.}
We describe our preprocessing procedure and present the other ten topics in \secref{sec:appendix_topics}.

\begin{table}[]
\centering
\small
\begin{tabular}{llr}
\toprule
\multicolumn{1}{c}{\textbf{LDA topic}} & \multicolumn{1}{c}{\textbf{top words}} & \multicolumn{1}{c}{\textbf{average topic weight}} \\ \toprule
\topicname{personal opinion} & \begin{tabular}[c]{@{}l@{}}opinion, fact, reason, agree, understand,\\ medium, argument, talking, making, decision\end{tabular} & 0.083   \\ 
\midrule
\topicname{game strategy}     &   \begin{tabular}[c]{@{}l@{}}defense, offense, defender, defensive, shooting,\\ offensive, shoot, open, guard, post\end{tabular}    &   0.082        \\ \midrule
\topicname{season prospects}      & \begin{tabular}[c]{@{}l@{}}final, playoff, series, won, championship,\\ beat, winning, west, east, title\end{tabular}     &   0.078         \\ \midrule
\topicname{future}       & \begin{tabular}[c]{@{}l@{}}pick, trade, star, top, chance,\\ young, future, move, round, potential \end{tabular}  &  0.075 \\ \midrule
\topicname{game stats}    & \begin{tabular}[c]{@{}l@{}}top, number, league, stats, mvp,\\ average, career, assist, put, shooting \end{tabular}             &   0.075  \\ \bottomrule
\end{tabular}
\caption{The top five topics by LDA using all the comments in \communityname{/r/NBA}. 
The top ten weighted words are presented for each topic. 
In preprocessing, all team and player names are removed. 
The remaining words are converted to lower case and lemmatized before training the LDA model.
}
\label{tab:topic}
\end{table}

\section{Research Questions and Hypotheses}

We study three research questions to understand how team performance affects fan behavior in online fan communities.
The first one is concerned with team performance in a single game and that game's associated user activity, while the other two questions are about team performance in a season and community properties (fan loyalty and the topics of discussion).

\subsection{Team Performance and Game-level Activity}

An important feature of NBA-related subreddits is to support game related discussion.
In practice, each game has a game thread in the home-team subreddit, the away-team subreddit, and the overall \communityname{/r/NBA}.
Team performance in each game can have a short-term impact on fans' behavior. 
For instance, \citet{Leung2017Effect} show that losing games has a negative impact on the contributions to the corresponding team's Wikipedia page, but winning games does not have a significant effect. 
However, it remains an open question how team performance in games relate to user activity in \textit{online sports fan communities}.

Previous studies find that fans react differently to 
top
teams than to bottom teams
based on interviews and surveys \cite{doyle2017there,sutton1997creating,yoshida2015predicting}. 
In particular, \citet{doyle2017there} find that fans of teams with an overwhelming loss to win ratio can be insensitive to losses through interviews.
In contrast, fans that support top teams may be used to winning. The hype created by the media and other fans elevates the expectation in the fan community. As a result, losing can surprise fans of the top teams and lead to a heated discussion. 
Therefore, we formulate our first hypothesis as follows:

\smallskip
\noindent\textbf{H1:} In subreddits of the top teams, fans are more active on losing days; in subreddits of the bottom teams, fans are more active on winning days.

\subsection{Team Performance and Fan Loyalty}

Researchers in sports management show that a team's recent success does not necessarily lead to a loyal fan base
\cite{campbell2004beyond,bee2010exploring,stevens2012influence}. 
For instance, 
``bandwagon fan'' refers to
individuals who become fans of a team simply because of their recent success. 
These fans tend to have a weak attachment to the team and are ready to switch to a different team when the team starts to perform poorly.
On the contrary, in the bottom teams, 
active fans that stay during adversity are probably loyal due to their deep attachment to the team
\cite{doyle2017there,campbell2004beyond}. 
They are able to endure current stumbles and treat them as a necessary process for future success. 
Our second hypothesis explores the relation between team performance and fan loyalty:

\smallskip
\noindent\textbf{H2:} Top team subreddits have lower fan loyalty and bottom team subreddits have higher fan loyalty.

\subsection{Team Performance and Topics of Discussion}

In addition to whether fans stay loyal, our final question examines what fans talk about in an online fan community.
As a popular sports quote says, 
``{\it Winning isn't everything, it's the only thing}.''\footnote{Usually attributed to UCLA football coach Henry Russel Sanders.}
A possible hypothesis is that that the discussion concentrates on winning and team success.
However, we recognize the diversity across teams depending on team performance.
Several studies find that fans of teams with poor performance may shift the focus from current failure to future success: staying optimistic can help fans endure adversity and maintain a positive group identity \cite{doyle2017there,campbell2004beyond,jones2000model}.
This is in clear contrast with the focus on winning the final championship of the top teams \cite{campbell2004beyond}. 
As a result, we formulate our third hypothesis as follows:

\smallskip
\noindent\textbf{H3:} The topics of discussion in team subreddits vary depending on team performance. 
Top team subreddits focus more on \topicname{season prospects}, while bottom team subreddits focus more on \topicname{future}.

\section{Method}

In this section, we first provide an overview of independent variables and then discuss dependent variables and formulate 
\replaced{hierarchical regression analyses}{linear models} to test our hypothesis in each research question.

\subsection{Independent Variables}
To understand how team performance affects fan behavior in online fan communities, we need to control for factors such as a team's market value and average player age. 
We collect statistics of the NBA teams from the following websites:
Fivethirtyeight,~\footnote{\url{http://fivethirtyeight.com/}.} 
Basketball-Reference,~\footnote{\url{https://www.basketball-reference.com/}.} 
Forbes.~\footnote{\url{https://www.forbes.com/}.} and Wikipedia.~\footnote{\url{https://www.wikipedia.org/}.}
We standardize all independent variables for linear regression models.
Table~\ref{tab:variables} provides a full list of all variables used in this paper. 
In addition to control variables that capture the differences between seasons and months, the variables can be grouped in three categories: performance, game information, and team information.

\begin{table}[t]
\centering
\small
\begin{tabular}{lp{0.5\textwidth}r}
\toprule
\multicolumn{1}{c}{\textbf{Variable}}              & \multicolumn{1}{c}{\textbf{Definition}}                                         & \multicolumn{1}{c}{\textbf{Source}} \\ \midrule
\multicolumn{3}{l}{\textit{Performance}}                                                   \\
{\bf winning}          & Win or lose a game.               				& FiveThirtyEight    \\
{\bf season elo}            & A team's elo rating at the end of a season.          & FiveThirtyEight    \\ 
{\bf season elo difference}            & A team's elo rating difference between the end of a season and its last season.          & FiveThirtyEight    \\ 
{\bf month elo}             & A team's elo rating at the end of that month.             & FiveThirtyEight \\
{\bf month elo difference}             & A team's elo rating difference between the end of a month and its last month.            & FiveThirtyEight    \\ 
\midrule
\multicolumn{3}{l}{\textit{Game information}}   \\
team elo              & A team's elo rating before the game.                   & FiveThirtyEight    \\
opponent elo          & The opponent's elo rating before the game.               & FiveThirtyEight    \\
point difference        & Absolute point difference of the game.                & FiveThirtyEight    \\ 
rivalry or not		& If the opponent team is a rivalry. 							& Wikipedia \\                                             
top team            &   If a team is with the five highest elo ratings at the end of a season.        & FiveThirtyEight      \\
bottom team         &   If a team is with the five lowest elo ratings at the end of a season.     & FiveThirtyEight                 \\
\midrule
\multicolumn{3}{l}{\textit{Team information}}                                                   \\
market value          & Transfer fee estimated to buy a team on the market.             & Forbes \\
average age           & The average age of the roster.              & Basketball-Reference    \\
\added{\#star players}     & The number of players selected to play the NBA All-Star Game.& Basketball-Reference  \\
\added{\#unique users}& The number of users that made at least one post/comment in the team's subreddit.   & N/A \\
offense       & The average points scored per game.                 & Basketball-Reference    \\
defense        & The average points allowed per game.                 & Basketball-Reference    \\ 
turnovers        & The average turnovers per game.                 & Basketball-Reference    \\  \midrule
\multicolumn{3}{l}{\textit{Temporal information}} \\
season & A categorical variable to indicate the season. & N/A (control variable) \\
month & A categorical variable to indicate the month. & N/A (control variable) \\
\bottomrule
\end{tabular}
\caption{List of variables and their corresponding definitions and sources. Measurements of team performance are in bold.
}
\label{tab:variables}
\end{table}

\subsubsection{Performance}

Since our research questions include both team performance in a single game and team performance over a season,
we consider performance variables both for a game and for a season.
First, to measure a team's game performance, we simply use whether this team wins or loses.
Second, to measure a team's performance over a season, we use elo ratings of the NBA teams.
The elo rating system was originally invented as 
a chess rating system for calculating the relative skill levels of players. 
The popular forecasting website FiveThirtyEight developed an elo rating system to 
measure the skill levels of different NBA teams~\cite{elo538}. 
These elo ratings are used to predict game results on FiveThirtyEight and are well received by major sports media, such as ESPN~\footnote{\url{http://www.espn.com/}.}
 and CBS Sports.~\footnote{\url{https://www.cbssports.com/}.} 
The FiveThirtyEight elo ratings satisfy the following properties:
\begin{itemize}
\item A team's elo rating is represented by a number that increases or decreases 
depending on the outcome of a game. After a game, the winning team takes 
elo points from the losing one, so the system is zero-sum.
\item The number of elo points exchanged after a game depends on the elo rating difference between two teams prior to the game, 
final basketball points, and home-court advantage. Teams gain more elo points for unexpected wins, great basketball point differences, and winning away games.
\item The long-term average elo rating of all the teams is 1500. 
\end{itemize}

To measure team performance, we use a team's elo rating at the end of a season as well as the elo difference between the end of this season and last season.
A high elo rating at the end indicates an absolute sense of strong performance;
a great elo rating difference suggests that a team has been improving.
In addition to studying how team performance over a season affects fan loyalty, we also include team performance over a month to check the robustness of the results.

\subsubsection{Game Information}
To test {\bf H1}, we need to use the interaction between game performance and top (bottom) team so that we can capture whether a top team loses or a bottom team wins.
We define {\em top team} as teams with the highest five elo ratings at the end of a season and {\em bottom team} as teams with the lowest five elo ratings at the end of a season.
We also include the following variables to characterize a single game:
1) Two team's elo ratings, which can partly measure the importance of a game;
2) (Basketball) point difference, which captures how close a game is; 
3) Rivalry game: which indicates known rivalry relations in the NBA, such as the Lakers and the Celtics.
We collect all pairs of the NBA rivalry teams from Wikipedia~\cite{wiki:rivalry}.

\subsubsection{Team Information}
To capture team properties, \replaced{we include a team's market value, average age of players, 
and the number of star players.}{we include both a team's market value and average age.}
Market value estimates the value of a team  on the current market. 
There are three key factors that impact a team's market value, 
including market size, recent performance and history~\cite{marketvalue}. 
We collect market values of all NBA teams from Forbes.
We scrape the average age of players on the roster from Basketball-Reference, 
which computes the average age of players at the start of Feb 1st of that season. 
The website chooses to calculate average age 
on Feb 1st because it is near the player trade deadline~\citep{NBARule}, and every team has a relatively stable roster at that time. 
\added{The number of star players measures the number of players selected 
to play in the NBA All-Star Game~\citep{wiki:NBA_All-Star_Game} of that season 
and this information is collected from Basketball-Reference.}
We further include variables that characterize a team's playing style: 
1) Offense: the average points scored per game;  
2) Defense: the average points allowed per game;
3) Turnovers: the average number of turnovers per game. 
\added{Teams' playing style information by season is collected from Basketball-Reference.}

\subsection{Analysis for \replaced{H1}{RQ1}}
\begin{figure}
    \begin{subfigure}[t]{0.48\textwidth}
        \includegraphics[width=\textwidth]{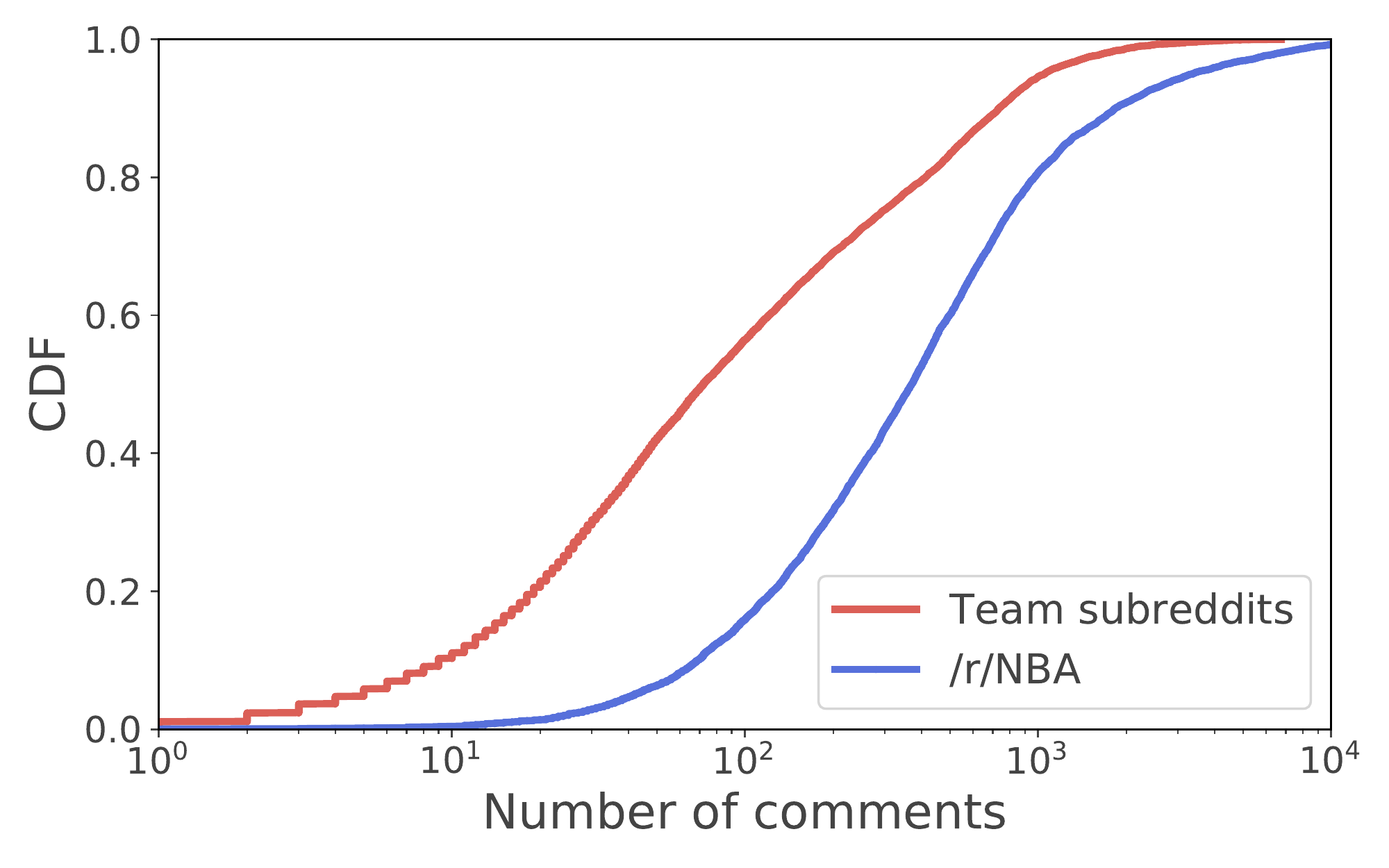}
        \caption{The CDF of \#comments in game threads.}
        \label{fig:CDFComment}
    \end{subfigure}
    \hfill
    \begin{subfigure}[t]{0.48\textwidth}
        \includegraphics[width=\textwidth]{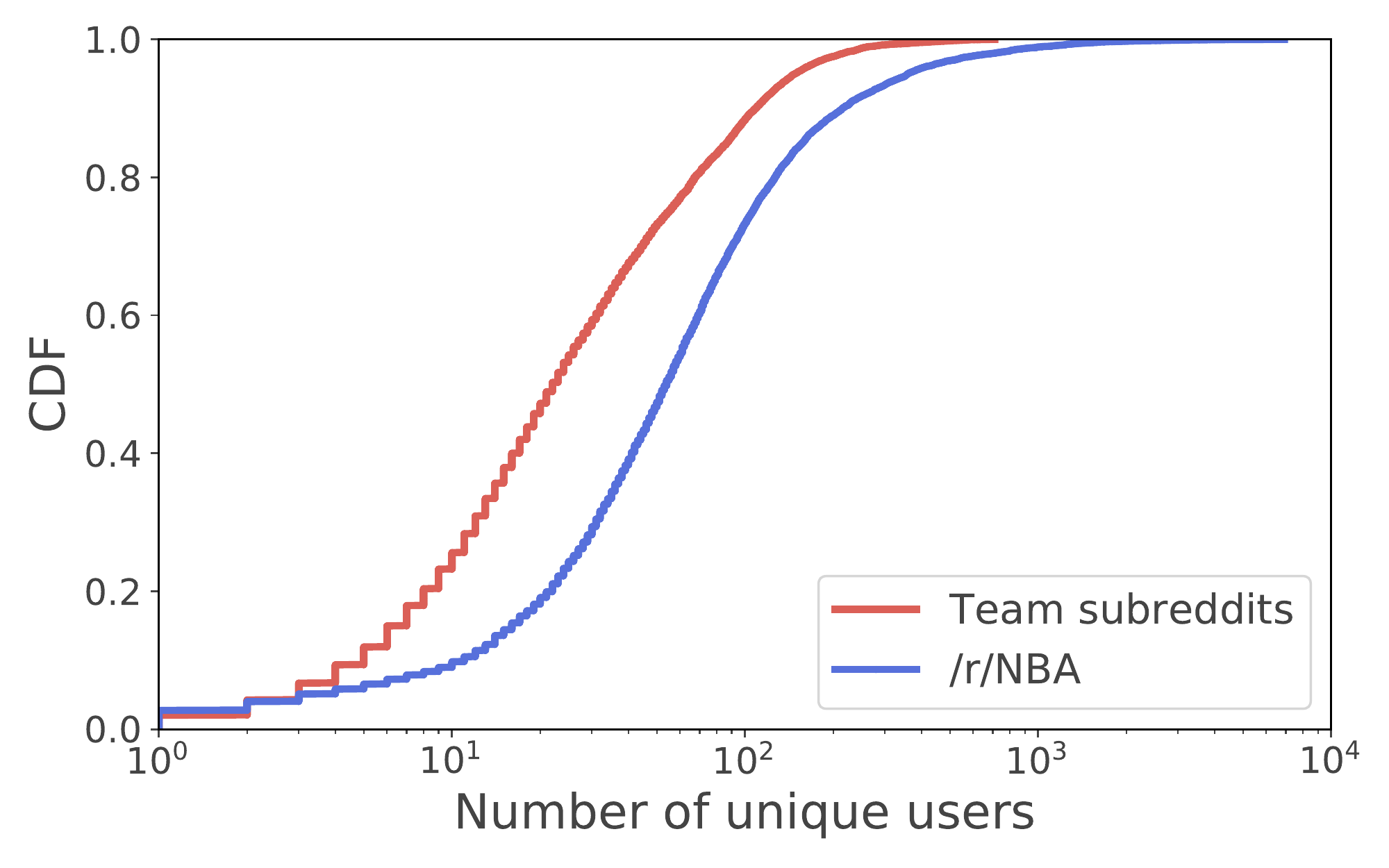}
        \caption{The CDF of \#unique users in game threads.}
        \label{fig:CDFUser}
    \end{subfigure}
    \caption{\added{Figure~\ref{fig:CDFComment} shows the cumulative distribution of the number
    of comments in all game threads in both \communityname{/r/NBA} and team subreddits.
    Figure~\ref{fig:CDFUser} shows the cumulative distribution of the number
    of unique users in all game threads in both \communityname{/r/NBA} and team subreddits.} 
    }
    \label{fig:gamethreadsCDF}
\end{figure}

\begin{figure}
\centering
\includegraphics[width=0.65\linewidth]{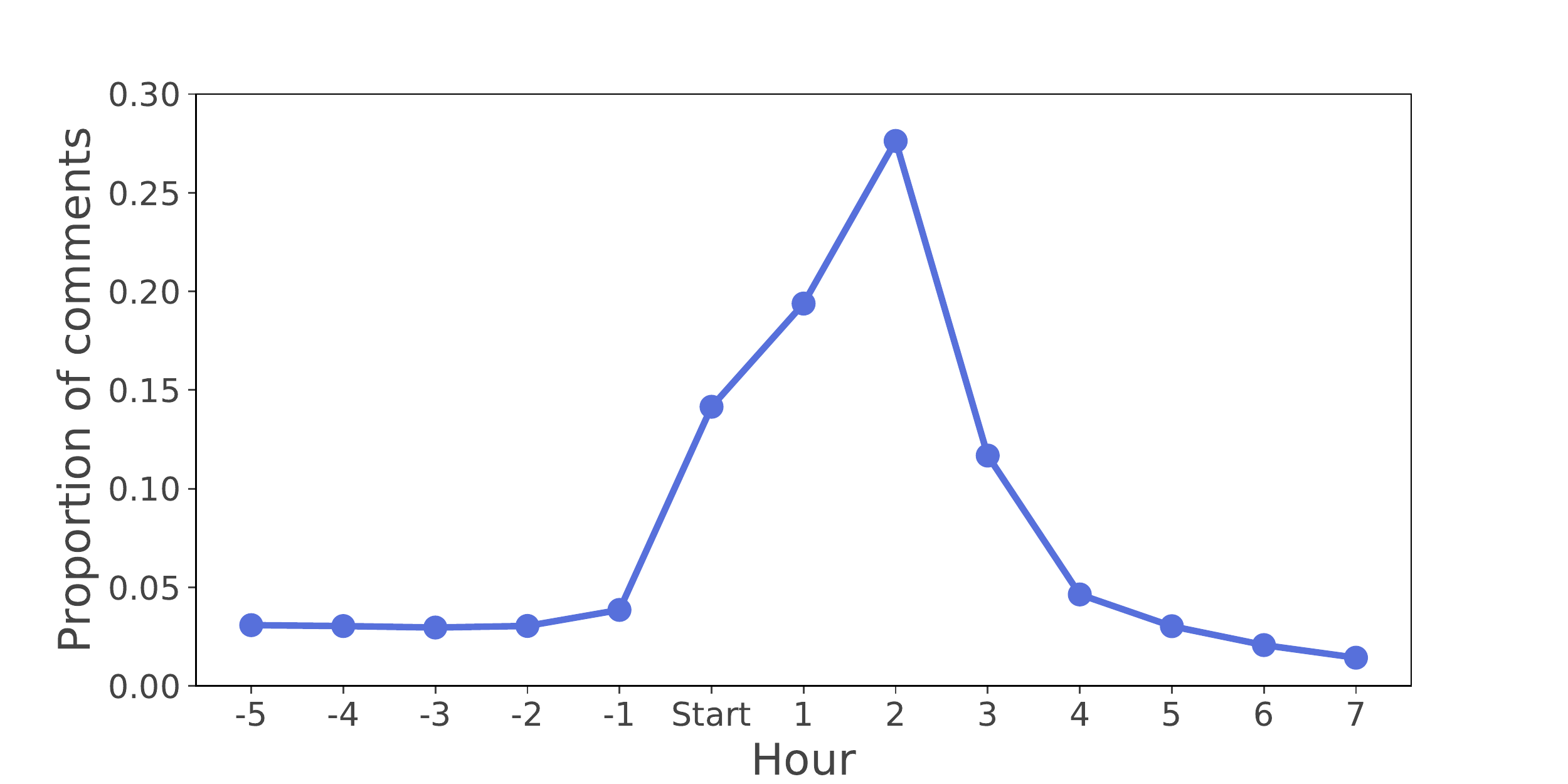}
\caption{%
\replaced{The average proportion of comments made in each team subreddit 
by hour on the game day during the 2017 season (normalized based on game's starting hour).
Error bars represent standard errors and are too small to see in the figure.
Comment activity increases and peaks at the second hour after the game starts, 
as a typical NBA game takes around 2.5 hours.}
{The distribution of comments by hour on two randomly picked consecutive game days 
in \communityname{/r/Lakers}. 
On 2017-01-12, the game started on 17:30
and the number of comments peaked
from 17:00 to 20:00, as a typical NBA game lasts around 2.5 hours. On 2017-01-14, the game started on 12:30, 
a similar peak occurred from 12:00 to 15:00.}
}
\label{fig:proportioncomments}
\end{figure}

Online fan communities provide a platform for fans to discuss sports games in real time and make the game watching experience interactive with other people on the Internet.
Accordingly, every team subreddit 
posts a ``Game Thread''
before the start of 
a game. 
Fans are encouraged to make comments related to a game in its game thread.
\added{Figure~\ref{fig:gamethreadsCDF} shows the cumulative distributions of the number of comments and the number of unique users.}
A game thread can 
accumulate hundreds or thousands of comments.
The number of comments is usually significantly higher during game time than other time periods. 
\replaced{Figure~\ref{fig:proportioncomments} shows the average proportion of comments 
made in each team subreddit by hour on the game day of the 2017 season 
(normalized based on games' starting hour). 
The number of comments peaked around the game time.}
{Figure~\ref{fig:proportioncomments} shows the proportion of comments 
by hour on two randomly picked consecutive 
game days in \communityname{/r/Lakers} and the number of comments 
peaked around game time.}

We use the number of comments in game threads to 
capture the fan activity level for a game.
Most game threads used \replaced{titles that are similar to this}{the title} format: ``[Game Thread]: team 1 @ team 2''.\footnote{If more than one game thread is created for the same game, 
only the first one is kept, and the others are deleted by the moderator.}
\replaced{We}{After a careful regular expression matching, we} detected 8,596 game threads 
in team subreddits and 6,277 game threads in \communityname{/r/NBA} \added{based on regular expression matching}.
Since NBA-related subreddits allow any fan to create game threads, 
titles of game threads
do not follow the same
pattern, especially in the earlier times of team subreddits.
A detailed explanation and sanity check is presented in \secref{sec:appendix_thread}.

\replaced{Hierarchical}{OLS} regression analysis was used to analyze the effect of team performance 
in a single game on fan activity.
Our \replaced{full}{formal} linear regression model to test {\bf H1} 
is shown below:
\begin{align}
\label{eq:h1}
\variablename{\#comments in game thread} \sim &  \beta_0 + \added{\beta_s\,\variablename{season} + \beta_m\,\variablename{month}   
+ \beta_t\,\variablename{top team} + \beta_b\,\variablename{bottom team}} \nonumber\\
& + \beta_1\,\variablename{winning} + \beta_2\,\variablename{top team winning} + \beta_3\,\variablename{top team losing} \nonumber \\
& + \beta_4\,\variablename{bottom team winning} + \beta_5\,\variablename{bottom team losing} \nonumber \\
& + \beta_6\,\variablename{team elo} + \beta_7\,\variablename{opponent elo} + \beta_{8}\,\variablename{rivalry or not} 
+ \beta_{9}\,\variablename{point difference}  \nonumber \\
& + \beta_{10}\,\variablename{market value} + \beta_{11}\,\variablename{average age} + \added{\beta_{12}\,\variablename{\#star players}} + \added{\beta_{13}\,\variablename{\#unique users}} \nonumber \\
& + \beta_{14}\,\variablename{offense} + \beta_{15}\,\variablename{defense} + \beta_{16}\,\variablename{turnovers} \nonumber.\\
\end{align}

To test our hypothesis in team subreddits, all the variables in Equation~\ref{eq:h1} are included.
Unlike game threads in team subreddits, game threads in \communityname{/r/NBA} %
involve two teams and the following variables are ill-defined: 
``winning,'' ``offense,'' ``defense,'' and ``turnovers.''
Therefore, these variables are removed when testing our hypothesis on game threads in \communityname{/r/NBA}. 

\subsection{Analysis for \replaced{H2}{RQ2}}
Fan loyalty refers to people displaying recurring behavior and a strong 
positive attitude towards a team~\citep{dwyer2011divided}. 
To examine the relationship between team performance and fan loyalty in team subreddits, we first define active users as those that
post or comment in a team subreddit during a time period.
We then define two measurements of fan loyalty: 
\textit{seasonly user retention} and \textit{monthly user retention}.
Seasonly user retention refers to the proportion of users that remain active in season $s+1$ among all users that are active in season $s$.
Monthly user retention refers to the proportion of users that remain active in month $m+1$ among all users that are active in month $m$.
The \replaced{full}{formal} linear regression models to test {\bf H2}
are shown below:
\begin{align}
\label{eq:h2season}
\variablename{seasonly user retention} \sim &  \beta_0 + \added{\beta_s\,\variablename{season}} \nonumber \\ 
& + \beta_1\,\variablename{season elo}  + \beta_2\,\variablename{season elo difference}  \nonumber \\
& + \beta_{3}\,\variablename{market value} + \beta_{4}\,\variablename{average age} + \added{\beta_{5}\,\variablename{\#star players}} + \added{\beta_{6}\,\variablename{\#unique players}} \nonumber \\ 
&+ \beta_{7}\,\variablename{offense} + \beta_{8}\,\variablename{defense} + \beta_{9}\,\variablename{turnovers} \nonumber. \\ 
\end{align}

\begin{align}
\label{eq:h2season}
\variablename{monthly user retention} \sim &  \beta_0 + \added{\beta_s\,\variablename{season} 
+ \beta_m\,\variablename{month}} \nonumber \\  
& + \beta_1\,\variablename{month elo}  + \beta_2\,\variablename{month elo difference}  \nonumber \\
& + \beta_{3}\,\variablename{market value} + \beta_{4}\,\variablename{average age} + \added{\beta_{5}\,\variablename{\#star players}} + \added{\beta_{6}\,\variablename{\#unique players}} \nonumber \\ 
&+ \beta_{7}\,\variablename{offense} + \beta_{8}\,\variablename{defense} + \beta_{9}\,\variablename{turnovers} \nonumber. \\ 
\end{align}

\subsection{Analysis for \replaced{H3}{RQ3}}
Among the five topics listed in Table~\ref{tab:topic}, \topicname{season prospects} and
\topicname{future}  topics are closely related to our hypotheses about fans talking about winning and framing the future.
By applying the trained LDA model to comments in each team subreddit, 
we are able to estimate the average topic distribution of each team subreddit by season. 
Our \replaced{full}{formal} linear regression model to test {\bf H3} 
is shown below:
\begin{align}
\label{eq:h3}
\variablename{topic weight} \sim &  \beta_0 + \added{\beta_s\,\variablename{season}} \nonumber \\  
& + \beta_1\,\variablename{season elo}  + \beta_2\,\variablename{season elo difference}  \nonumber \\
& + \beta_{3}\,\variablename{market value} + \beta_{4}\,\variablename{average age} + \added{\beta_{5}\,\variablename{\#star players}} + \added{\beta_{6}\,\variablename{\#unique players}} \nonumber \\ 
&+ \beta_{7}\,\variablename{offense} + \beta_{8}\,\variablename{defense} + \beta_{9}\,\variablename{turnovers} \nonumber, \\  
\end{align}
where \variablename{topic weight} can be the average topic weight of either  \topicname{season prospects} or \topicname{future.}

\begin{table}[t]
\centering
\small
\begin{tabular}{l|LLL|LLL}
\toprule
                              & \multicolumn{3}{c|}{Team subreddits}                                             & \multicolumn{3}{c}{\communityname{/r/NBA}}                                                     \\
\multicolumn{1}{c|}{Variable} & \multicolumn{1}{c}{Reg. 1} & \multicolumn{1}{c}{Reg. 2} & \multicolumn{1}{c|}{Reg. 3} & \multicolumn{1}{c}{Reg. 1} & \multicolumn{1}{c}{Reg. 2} & \multicolumn{1}{c}{Reg. 3} \\
\midrule
\textit{Control: season} &&&&&& \\
2014				& 0.011***	& 0.012***	& 0.013***	& 0.007***	& 0.007***	& 0.005***	\\
2015				& 0.032***	& 0.032***	& 0.045***	& 0.010***	& 0.010***	& 0.006***	\\
2016				& 0.046***	& 0.046***	& 0.062***	& 0.014***	& 0.015***	& 0.012***	\\
2017				& 0.067***	& 0.068***	& 0.081***	& 0.018***	& 0.019***	& 0.016***	\\ [5pt]
\textit{Control: top/bottom team} &&&&&& \\
top team			& 			& 0.012***	& 0.012***	& 			& 0.012***	& 0.020*** 	\\
bottom team			& 			& -0.012***	& -0.007***	&			& -0.009***	& -0.007**	\\ [5pt]
\textit{Performance} &&&&&& \\
winning 			&           &  			& 0.003**  	& 			& 			& \multicolumn{1}{c}{--}  \\
top team winning    &           &     		& -0.006*** &           &    		& -0.018***  \\
top team losing     &           &       	& 0.006***  &     		&      		& 0.018***   \\
bottom team winning &           &  			& 0.007***  &      		&       	& 0.006***   \\
bottom team losing  &  			&     		& -0.011*** &  			&       	& -0.003*    \\ [5pt]
\textit{Game information} &&&&&& \\
team elo            &   	 	&         	& 0.083***  &     		&       	& 0.091***   \\
opponent elo        &           &  			& 0.070***  &  			&   		& 0.111***   \\
rivalry or not      &           &      		& 0.010***  &   		&    		& 0.008***   \\
point difference    &           &           & -0.017*** &  			&  			& -0.010***  \\ [5pt]
\textit{Team information}  &&&&&&\\
market value        &           &    		& 0.051***  &   		&    		& 0.013***  \\
average age         &           &    		& -0.067*** &   		&   		& -0.015*  \\
\added{\#star players}	&			&			& 0.040***	&			&			& 0.020***	\\
\added{\#unique users}&         &           & 0.058***  &           &           & 0.017***  \\
offense             &           &    		& 0.023**  & 			& 			& \multicolumn{1}{c}{--}\\
defense             &           &   		& -0.012** & 			& 			& \multicolumn{1}{c}{--} \\
turnovers           &           &    		& 0.084***  & 			& 			& \multicolumn{1}{c}{--} \\
\midrule
intercept           & -0.010**  & -0.011**   & 0.085***    & 0.001      & -0.004**      & -0.156***     \\
Adjusted $R^2$      & 0.236  	& 0.286   	& 0.440     & 0.302      & 0.338      & 0.644     \\
Intraclass Correlation (Season) \cite{packageICC} & 0.087  	& \multicolumn{1}{c}{--}   	& \multicolumn{1}{c|}{--}     & 0.021      & \multicolumn{1}{c}{--}      & \multicolumn{1}{c}{--}     \\
\bottomrule
\end{tabular}
\caption{\replaced{Hierarchical regression analyses}{Linear regression models} for game-level activity 
in team subreddits and \communityname{/r/NBA}. 
\added{Month is also added as a control variable for each model.}
\textbf{Throughout this paper, the number of stars indicate p-values, ***: 
$p<0.001$, **: $p<0.01$ *: $p<0.05$.}
\added{We report $p$-values without the Bonferroni correction in all the regression tables.
In \secref{sec:ftest}, we report $F$-test results with the null hypothesis that adding team performance variables does not provide a significantly better fit and reject the null hypothesis after the Bonferroni correction.}
}
\label{tab:single}
\end{table}

\section{Results}

Based on the above variables, our results from \replaced{hierarchical regression analyses}
{linear regression models} by and large validate our hypotheses.
Furthermore, we find that the average age of players on the roster consistently plays an important role in fan behavior, 
while it is not the case for market value \added{and the number of star players}.

\subsection{How does Team Performance Affect Game-level Activity? (\replaced{H1}{RQ1})}

Consistent with {\bf H1}, regression results show that the
top team losing and the bottom team winning correlate with higher levels of fan activity in both team subreddits and \communityname{/r/NBA}. 
Table~\ref{tab:single} presents the results of our \replaced{hierarchical regression analyses}{OLS linear regression}.
The $R^2$ value is \replaced{0.40}{0.39} for team subreddits and \replaced{0.63}{0.52} for \communityname{/r/NBA}, 
suggesting that our \replaced{linear variables}{linear regression} can reasonably \replaced{recover}{predict} fan 
activity
in game threads. 
Overall, fans are more active when their team wins in team subreddits 
(remember that the notion of one's team does not hold in \communityname{/r/NBA}).
The interaction with the top team and the bottom team show that surprise can stimulate fan activity: 
both the top team losing and the bottom team winning have significantly positive coefficients.
To put this into context, in the 2017 season, the average winning percentage of the top five teams is 69\%. 
Fans of the top teams may get used to their teams winning games, in which case losing becomes a surprise. 
On the contrary, the average winning percentage of the bottom five teams is 31\%. 
It is \replaced{invigorating}{``surprising''} for these fans to watch their team winning. 
The \replaced{extra excitement}{surprise} can stimulate more comments in the game threads in both team subreddits and \communityname{/r/NBA}. 
In comparison, when top teams win or bottom teams lose, fans are less active, evidenced by the negative coefficient in team subreddits (not as statistically significant in \communityname{/r/NBA}).

\begin{figure}[t]
\centering
\includegraphics[width=\textwidth]{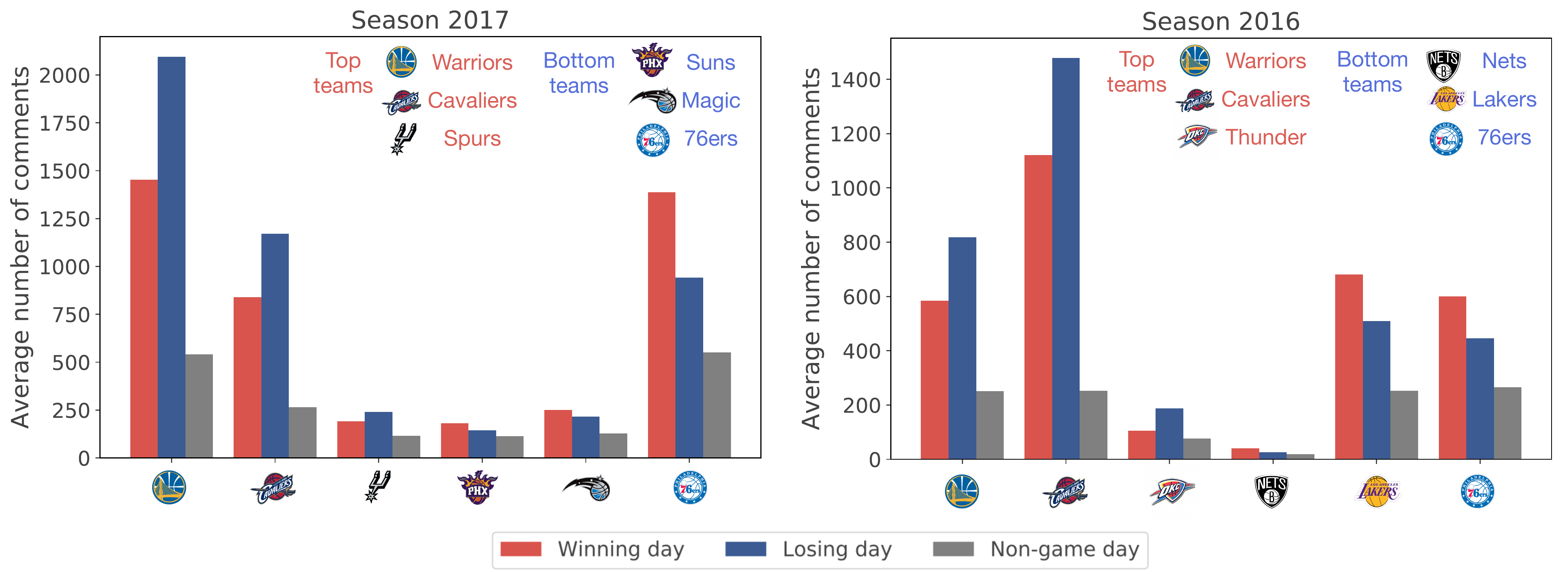}  
\caption{Average number of comments on winning, losing and non-game days 
for the top three and the bottom three teams in the 2017 (left) and 2016 (right) regular season. 
In all the top and bottom teams, 
the average number of comments on game days is significantly higher than non-game days. 
In all top teams, the average number of comments on losing days is higher than winning days, 
while bottom teams show the opposite trend.
}
\label{fig:commentvolume}
\end{figure} 

To further illustrate this contrast,
Figure~\ref{fig:commentvolume} 
shows the average number of comments 
on winning, losing, and non-game days for the top three and the bottom three teams in the 2017 and 2016 regular season. 
Consistent patterns arise: 1) In all top and bottom teams, 
the average number of comments on game days is significantly higher than non-game days;
2) In all top teams, the average number of comments on losing days is higher than winning days, but bottom teams show exactly the opposite trend.
Our results differ from that of \citet{Leung2017Effect}\replaced{, which finds that}{where} unexpected winning does not have a significant impact on Wikipedia page edits.
\added{One of the primary differences between our method and theirs is that 
they did not specifically control the effect of top/bottom team.}
It \replaced{may also}{can} be explained by the fact that Wikipedia page edits 
do not capture the behavior of most fans and are much more sparse than comments in online fan communities.
Online fan communities provide rich behavioral data 
for understanding how team performance affects fan behavior.
\added{The number of fans involved in our dataset is much higher than that in their Wikipedia dataset. 
A comparison between fans' behavior on Reddit and Wikipedia could be an interesting
direction for future research.}
In addition, game information and team information also serve as important \replaced{factors}{predictors}. 
Among variables about game information, point difference is negatively correlated with game-level user activity, as
the game intensity tends to be higher when the point difference is small (a close game).
Better teams (with higher elo ratings) 
playing against better teams or rivalry teams correlates with higher user activity levels.
As for team information, a team's market value\replaced{, the number of unique users, and the number of star players are}{ is} 
positively correlated with the number of comments, 
\replaced{since these two factors are}{since market value is} 
closely related to the number of fans. 
Younger teams with more average points scored and 
less points allowed per game stimulate more discussion in team subreddits.

\begin{table}[t]
\small
\centering
\begin{tabular}{l|LL|LL}
\toprule
\multicolumn{1}{l|}{}  & \multicolumn{2}{c|}{Seasonly user retention}  & \multicolumn{2}{c}{Monthly user retention} \\
\multicolumn{1}{c|}{Variable}   & \multicolumn{1}{c}{Reg. 1}  & \multicolumn{1}{c|}{Reg. 2}  
		& \multicolumn{1}{c}{Reg. 1}                 & \multicolumn{1}{c}{Reg. 2}         \\ \midrule
\textit{Control: season}  &&&&\\
2014        			& 0.237*** 	& 0.237***  				& 0.086*** 	& 0.055***					\\
2015   					& 0.184***  & 0.187***  				& 0.126*** 	& 0.126***					\\
2016       				& 0.088***  & 0.116*** 					& 0.130***	& 0.139***					\\
2017      				& 0.073***  & 0.090***  				& 0.137*** 	& 0.141***					\\[5pt]
\textit{Performance}  	&&&&\\
season elo         		& 			& -0.370**					& 		 	& \multicolumn{1}{c}{--}	\\
season elo difference   & 			& 0.229***  				& 		 	& \multicolumn{1}{c}{--}	\\
month elo       		&  			& \multicolumn{1}{c}{--}	& 			& -0.170**					\\
month elo difference    &           & \multicolumn{1}{c}{--}   	&   		& 0.032**					\\[5pt]
\textit{Team information} &&&&\\
market value      		&			& 0.068*   					& 			& 0.051**					\\
average age     		&			& -0.105*					& 			& -0.021					\\
\added{\#star players}  &			& -0.038					&   		& 0.041		\\
\added{\#unique users}  &           & 0.181***                  &           & 0.111***    \\
offense  				&			& -0.168					&			& 0.004						\\
defense  				&			& -0.077					& 			& -0.053 					\\
turnovers  				&			& -0.037					&   		& -0.041					\\
\midrule
intercept        		& 0.583***	& 0.629***  				& 0.478***	& 0.460*** 					\\  
Adjusted $R^2$          & 0.286		& 0.503						& 0.155		& 0.232						\\
Intraclass Correlation (Season) \cite{packageICC}         & 0.396		& \multicolumn{1}{c|}{--}						& 0.029		& \multicolumn{1}{c}{--}						\\
\bottomrule
\end{tabular}
\caption{\replaced{Hierarchical regression analyses}{Linear regression models} 
for seasonly user retention rate and monthly user retention rate in team subreddits. 
\added{Month is also added as a control variable for the monthly user retention analysis.
{\em \#unique users} is counted every season for the seasonly user retention analysis 
and every month for the monthly user retention analysis.}
For both dependent variables, team's overall performance has a negative 
coefficient while short-term performance and 
market value has a positive coefficient. 
\deleted{Average age and playing style don't have significant effects 
on user retention rate.}
}
\label{tab:loyalty}
\end{table}

\begin{figure}[t]
\centering
  \includegraphics[width=\textwidth]{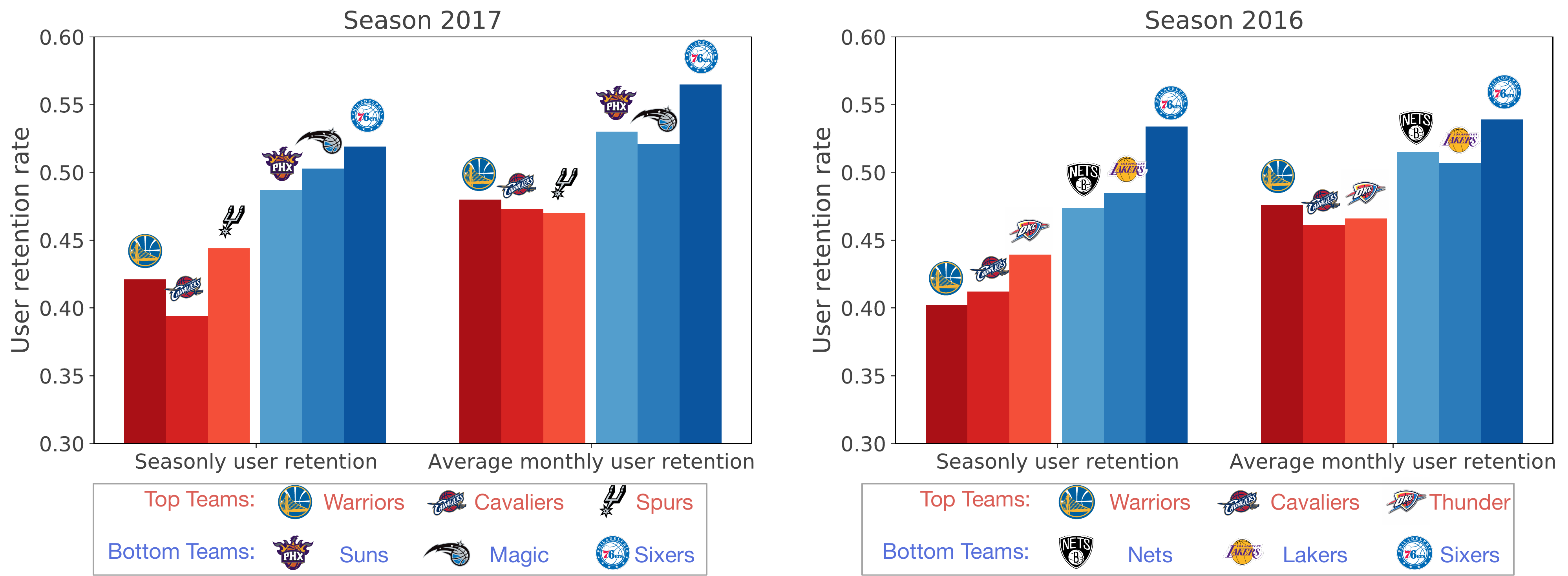}  
\caption{Seasonly user retention rate and average monthly user retention rate of the top three and bottom three teams in the 2017 (left) and 2016 (right) season. 
Bottom teams consistently have higher user retention than top teams.}
\label{fig:retention}
\end{figure}

\subsection{How does Team Performance Relate to Fan Loyalty in Team Subreddits? (\replaced{H2}{RQ2}) }

Our findings confirm \textbf{H2}, that top teams tend to have lower fan loyalty
and bottom teams tend to have higher fan loyalty, measured by both seasonly user retention and monthly user retention. 
Table~\ref{tab:loyalty} shows the \replaced{hierarchical}{linear} regression results.
In both regression \replaced{analyses}{models}, elo rating, 
which measures a team's absolute performance, 
has a statistically significant negative impact on user retention rate.
The coefficient of elo rating also has the greatest absolute value among all variables (except intercept).
Meanwhile, improved performance reflected by elo difference positively correlates with user retention.

Figure~\ref{fig:retention} presents the seasonly user retention rate 
and average monthly user retention rate of the top 3 and bottom 3 teams in the 2017 (left) and 2016 (right) season. 
It is consistent that in these two seasons, 
bottom teams have higher user retention rate than top teams, both seasonly and monthly.
This may be explained by the famous ``bandwagon'' phenomenon in professional sports~\cite{wann1990hard}: 
Fans may ``jump on the bandwagon'' by starting to follow the current top teams,
which provides a short cut to achievement and success for them.
In comparison, terrible team performance can serve as a loyalty filter. 
After a period of poor performance, 
only die-hard fans 
stay active and optimistic in the team subreddits. 
Our results echo the finding by \citet{hirt1992costs}:
after developing strong allegiances with a sports team, fans find it difficult 
to disassociate from the team, even when the team is unsuccessful.
It is worth noting that the low fan loyalty of the top teams cannot simply be explained by the fact that they tend to have more fans.
\deleted{Although we did not explicitly include the number of users as an independent variable because market value is highly correlated with the number of users (Pearson correlation at 0.71),
our results are robust, even if we include \#users in the regression models.}
In fact, teams with higher market value \added{and more unique users} (more fans) tend to have a higher user retention rate, partly because their success depends on a healthy and strong fan community.

Similar to game-level activity, fans are more loyal to younger teams, at least in seasonly user retention \added{(the coefficient is also negative for monthly user retention and $p-$value is 0.07)}.
Surprisingly, according to our \added{hierarchical} regression results, a
team's \added{number of star players and} playing style (offense, defense, and turnovers) 
\replaced{have}{has} no significant impact on user retention.

\begin{table}[t]
\small
\centering
\begin{tabular}{l|LL|LL}
\toprule
\multicolumn{1}{l|}{}  & \multicolumn{2}{c|}{\topicname{season prospects}}  & \multicolumn{2}{c}{\topicname{future}} \\
\multicolumn{1}{c|}{Variable}   & \multicolumn{1}{c}{Reg. 1}  & \multicolumn{1}{c|}{Reg. 2}  
		& \multicolumn{1}{c}{Reg. 1}                 & \multicolumn{1}{c}{Reg. 2}         \\ \midrule
\textit{Control: season}  &&&&\\
2014        			& 0.096*** 	& 0.056**  & 0.059* 	& 0.137***	\\
2015   					& 0.067**  	& 0.049*  	& 0.054* 	& 0.129***	\\
2016       				& 0.069**  & 0.053* 	& 0.109***	& 0.175***	\\
2017      				& 0.061*  	& 0.048*  	& 0.065* 	& 0.160***	\\[5pt]
\textit{Performance}  	&&&&\\
season elo         		& 			& 0.410***	& 		 	& -0.415**	\\
season elo difference   & 			& -0.130  	& 		 	& -0.099	\\[5pt]
\textit{Team information} &&&&\\
market value      		&			& -0.065  	& 			& 0.051	\\
average age     		&			& 0.149***	& 			& -0.189***	\\
\added{\#star players}		&			& 0.018		&			& -0.187**\\
\added{\#unique users}  &           & 0.071     &           & -0.104    \\
offense  				&			& -0.036	&			& 0.037		\\
defense  				&			& -0.120	& 			& -0.115 	\\
turnovers  				&			& 0.059		&   		& -0.092	\\ \midrule
intercept        		& 0.398***	& 0.286***  & 0.478***	& 0.814*** 	\\  
Adjusted $R^2$          & 0.013		& 0.578		& 0.002		& 0.619		\\
Intraclass Correlation (Season) \cite{packageICC} & 0.059		& \multicolumn{1}{c|}{--} & 0.003	& \multicolumn{1}{c}{--}		\\
\bottomrule
\end{tabular}
\caption{\replaced{Hierarchical regression analyses}{Linear egression models} for \topicname{season prospects} topic weight and \topicname{future} topic weight in team subreddits. 
Team performance has positive correlation 
with \topicname{season prospects} topic and negative correlation with \topicname{future} topic. }
\label{tab:rq2topic}
\end{table}
\begin{figure}[t]
\centering
  \includegraphics[width=\textwidth]{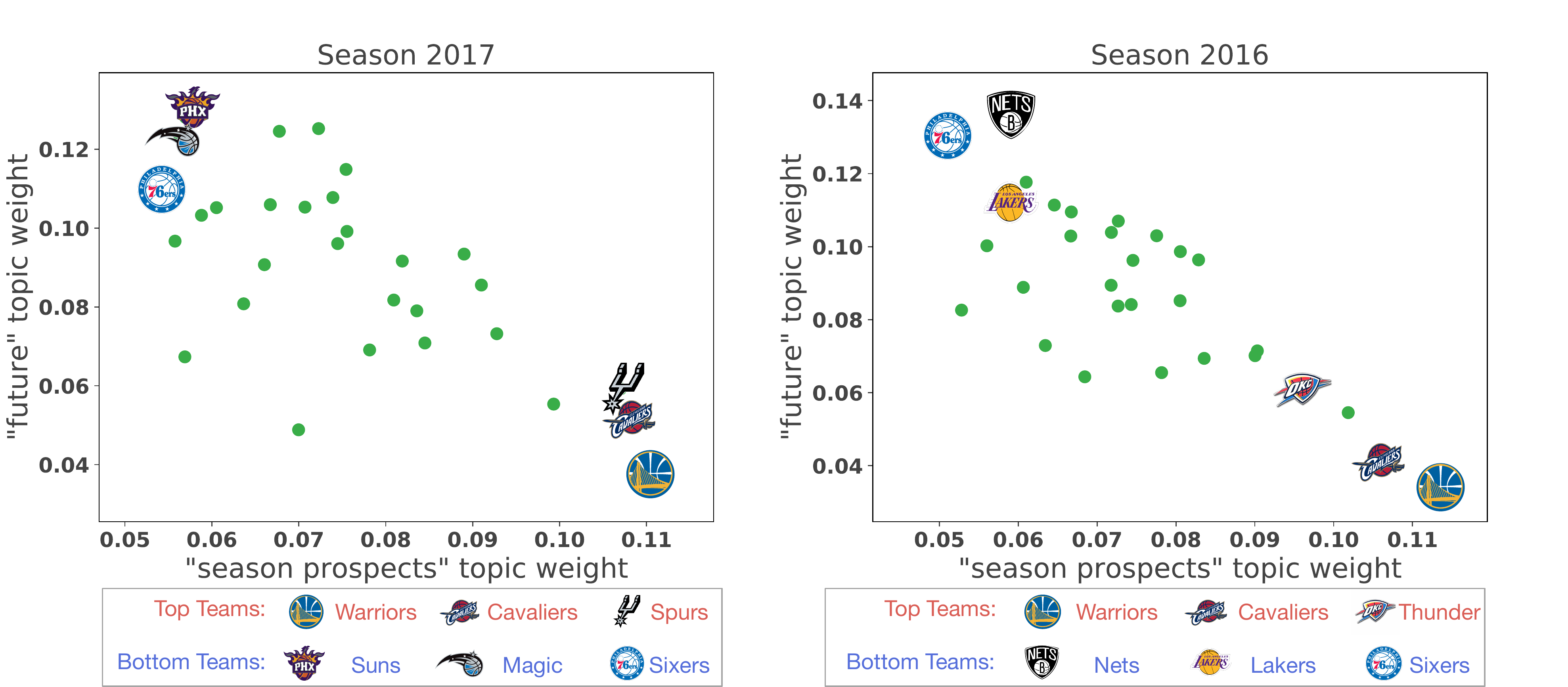}
\caption{Scatterplot of \topicname{season prospects} topic weight and \topicname{future} topic weight in all the team subreddits in the 2017 (left) and 2016 (right) season. 
The top three teams and bottom three teams are represented by team logos instead of points.
Teams are ranked by elo rating at the end of each season. Fans of the top teams tend to discuss much more \topicname{season prospects} topics (lower right corner) and
fans of the bottom teams tend to discuss much more \topicname{future} topics (upper left corner).
}
\label{fig:topic}
\end{figure}

\subsection{How does Team Performance Affect Topics of Discussion in Team Subreddits? (\replaced{H3}{RQ3})}
Our final question is concerned with the relation between team performance and topics of discussion in online fan communities.
Our results validate \textbf{H3}, that better teams have more discussions on \topicname{season prospects} and worse teams tend to discuss \topicname{future.}
Table~\ref{tab:rq2topic} presents the results of \replaced{hierarchical regression analyses}
{OLS linear regression} on \topicname{future} topic weight 
and \topicname{season prospects} topic weight computed with our LDA model. 
In both regressions, only team performance (season elo) and average age
have statistically significant coefficients.
Both team performance and average age are positively correlated with \topicname{season prospects} and negatively correlated with \topicname{future}.
\replaced{Moreover, the number of star players has a negative correlation with \topicname{future}
but has no significant effect on \topicname{season prospects.}}{Despite having only two variables with significant coefficients, both regression models achieve predictive power with $R^2$ above 0.55.}
\added{Despite having only two or three variables (except control variables and intercept) with significant coefficients, 
both regression analyses achieve strong correlation with $R^2$ above 0.57.}
Note that the improvement in team performance (season elo difference) does not have a significant effect.

As an example, Figure~\ref{fig:topic} further shows topic weights of \topicname{future} and \topicname{season prospects}
for all the teams in the 2017 and 2016 season. 
The top 3 teams and bottom 3 teams in each season are highlighted using team logos. 
The top teams are consistently in the lower right corner (high \topicname{season prospects}, low \topicname{future}),
while the bottom teams are 
in the upper left corner (low \topicname{season prospects}, high \topicname{future}). 
Our results echo the finding in \citet{doyle2017there}:
 framing the future is an important strategy for fans of teams with poor performance to maintain a positive identity 
in the absence of success.
The effect of average age reflects the promise that young talents hold for NBA teams.
Although it takes time for talented rookies that just come out of college to develop physical and mental strength to compete in the NBA,
fans can see great potential in them and 
remain positive about their team's future, despite the team's short-term poor performance. 
In contrast, veteran players are expected to bring immediate benefits to the team and compete for playoff positions and even championships.
For example, \citet{agingveteran} lists a number of veteran players 
who either took a pay cut or accepted a smaller role in top teams 
to chase a championship ring at the end of their career. 

A team's playing style, including offense points, defense points, and turnovers, doesn't seem to influence the topic weights of these two topics.
We also run regression for the other three top topics in Table~\ref{tab:topic} and present the results in \secref{sec:appendix_topic_regression}.
Team performance plays a limited role for the other three topics, while average age is consistently significant for all three discussion topics.

\section{Concluding Discussion}
\label{sec:conclusion}
In this work, we provide the first large-scale characterization
of online fan communities of the NBA teams.
We build a unique dataset that combines user behavior in 
NBA-related subreddits and statistics of team performance.
We demonstrate how team performance affects fan behavior both 
at the game level and at the season level.
Fans are more active when top teams lose and bottom teams win, 
which suggests that 
in addition to simply winning or losing,
surprise plays an important role in driving fan activity.
Furthermore, 
a team's strong performance doesn't necessarily make the fan community more loyal. 
It may attract ``bandwagon fans'' and result in a low user retention rate.
We find that the bottom teams generally have higher 
user retention rate than the top teams. 
Finally, fans of the top teams and 
bottom teams 
focus on different topics of discussion.
Fans of the top teams talk more about season records, playoff seeds, and winning
the championship, while fans of the bottom teams spend more time framing
the future to compensate for the lack of recent success.

\para{Limitations.} 
One key limitation of our work is the representativeness of our dataset.
First, although our study uses a dataset that spans five years, our period coincides with the rapid growth of the entire Reddit community.
We use \textit{season} and \textit{month} to try our best to account for temporal differences, 
but our sample could still be based upon fans with a mindset of growth.
Second, although \citet{rnba} suggests that \communityname{/r/NBA} is now playing an important role among fans,
the NBA fan communities on Reddit may not be representative of the Internet and the whole offline population.

Another limitation of our work lies in our measurement.
For game-level activity, we only consider the number of comments in the game threads.
This measurement provides a nice way to make sure that the comments are about the game, but we may have missed related comments in other threads.
We do not consider other aspects of the comments such as sentiment and passion.
In addition, our fan loyalty metric is entirely based on user retention.
A user who posts on a team subreddit certainly supports the team to a different extent from those who do not.
Our metric may fail to capture lurkers who silently support their teams.
Finally, our topics of discussion are derived from topic modeling, an unsupervised approach.
Supervised approaches could provide more accurate identification of topics, although the deduction approach would limit us to a specific set of topics independent of the dataset.

\para{Implications for online communities.}
First, our work clearly demonstrates that online communities do not only exist in the virtual world; they are usually 
embedded in the offline context and attract people with similar offline interests.
It is an important research question to understand to what extent and how online communities relate to offline contexts as well as what fraction of online communities are entirely virtual.
Professional sports provide an interesting case, 
because these online fan communities, in a way, only exist as a result of the offline sports teams and games.
Such connections highlight the necessity to combine multiple data sources to understand 
how fans' usage of social media correlates with the on-going events of the topic of their interests.
Our study has the potential to serve as a window into the relationship between online social behavior
and offline professional sports. We show that subreddit activity has significant correlations 
with game results and team properties. 
Exploring the factors that motivate users of interest-based communities to communicate with social media 
is also an important and rich area for future research. 
\added{For example, a promising future direction is to study the reasons 
behind fans departing a team subreddit.
Possible reasons include being disappointed
by the team performance or playing style, favorite players being traded, and being attacked by other fans in the team subreddit or \communityname{/r/NBA}.}

Second, our results show that teams with strong performance correlate with low fan loyalty.
These results relate to the multi-community perspective in online community research \cite{tan2015all,zhang2017community,hamilton2017loyalty,Zhu:2014:SEN:2556288.2557348}.
One future direction is to examine where fans migrate to and whether fans leave the NBA or the Reddit altogether, and more importantly, what factors determine such migration decisions.

Third, our findings reveal strategies for the design of sports-related online platforms. 
Our results clearly demonstrate that teams in under-performing periods 
are more likely to develop a more loyal fan base 
that discusses more about their team's \topicname{future.}
Recognizing these loyal fans and acknowledging their contributions within the fan community can be critical for
facilitating attraction and retention of these fans. 
For example, team subreddits' moderators may reward a unique flair to the users 
who have been active in the community for a long time, especially during the difficult times. 

\para{Implications for sports management.} 
Our findings suggest that winning is not everything.
In fact, unexpected losses can stimulate fan activity.
The increase of fan activity does not necessarily happen in a good way.
For example, the fans of the Cavaliers, which won the Eastern Championship of the 2017 season, started to discuss firing the team's head coach Tyronn Lue after losing three of the first six games in the following season.
Managers may try to understand the role of expectation in fan behavior and guide the increased activity and attention towards improving the team and building a strong fan base.

We also find that the average age of the roster consistently plays an important role in fan behavior:
younger teams tend to bring more fan activity on game days and develop a more loyal fan base that discusses about \topicname{future.}
These results contribute to existing literature on the effect of age in sports management.
\citet{timmerman2000racial} finds that the average age is positively correlated with team performance, while the age diversity is negatively correlated (in other words, veterans improve team performance but are not necessarily compatible with young players).
The tradeoff between veteran players and young talents requires more research from the perspective of both team performance and fan engagement.

Finally, it is crucial for teams to maintain a strong fan base that can support them during unsuccessful times because it is difficult for sports teams to sustain winning for a long time.
This is especially true in the NBA since the draft lottery mechanism is designed 
to give bottom teams opportunities to improve and compete.
Consistent with \citet{doyle2017there}, we find that framing the \topicname{future} can be an important strategy for teams with poor performance to maintain a positive group identity.
The absence of success can be a great opportunity to develop a deep attachment with loyal fans. Prior studies show that certain fan group would like 
to persevere with their supported team through almost anything, 
including years of defeat, to recognize themselves as die-hard fans. 
By doing this, they feel that they would reap more affective significance among the fan community 
when the team becomes successful in the future~\cite{wann1990hard,hyatt2015using}. 
It is important for managers to recognize these loyal fans and create ways to acknowledge and leverage their positions within the fan community.
For instance,  teams may host ``Open Day'' and invite these loyal fans to visit facilities and 
interact with star players and coaching staff. 
Hosting Ask Me Anything (AMA)~\cite{wiki:AMA} interviews is another strategy to engage with online fan communities.

\bibliographystyle{ACM-Reference-Format}
\bibliography{paper}

\appendix

\section{Appendix}

\subsection{Topic Modeling and Additional Topics}
\label{sec:appendix_topics}
The following pre-processing procedures are used to clean data before training topic models:
\begin{itemize}
\item Converting all the words to lower case.
\item Removing all the HTML links in the comments.
\item Removing all the player names and nicknames, such as Lebron, Kobe.
\item Removing all the team names, such as Lakers, Celtics.
\item Removing common stopwords. 
\item Lemmatizing all words.
\end{itemize}

The remaining 10 topics generated after training are shown in Table~\ref{tab:tentopic}.

\begin{table}[t]
\centering
\small
\begin{tabular}{llr}
\toprule
\multicolumn{1}{c}{\textbf{LDA topic}} & \multicolumn{1}{c}{\textbf{top words}} & \multicolumn{1}{c}{\textbf{average topic weight}} \\ \toprule
Topic 00 & \begin{tabular}[c]{@{}l@{}}big, level, work, hard, league,\\ talent, basketball, long, skill, playing\end{tabular} & 0.071   \\ \midrule
Topic 01     &   \begin{tabular}[c]{@{}l@{}}fucking, dude, day, kid, life,\\ friend, face, talking, bitch, court\end{tabular}    &   0.068  \\ \midrule
Topic 02      & \begin{tabular}[c]{@{}l@{}}injury, minute, played, playing, half,\\ quarter, end, night, ago, start\end{tabular}     &   0.068         \\ \midrule
Topic 03       & \begin{tabular}[c]{@{}l@{}}watch, watching, basketball, feel, fun, damn,\\ hope, god, fucking, honestly, suck \end{tabular}  &  0.067 \\ \midrule
Topic 04       & \begin{tabular}[c]{@{}l@{}}call, foul, ref, throw, free,\\ called, hand, rule, hit, foot \end{tabular}  &  0.065 \\ \midrule
Topic 05       & \begin{tabular}[c]{@{}l@{}}contract, money, deal, cap, million,\\ sign, free, pay, max, salary \end{tabular}  &  0.063 \\ \midrule
Topic 06       & \begin{tabular}[c]{@{}l@{}}post, comment, thread, edit, read,\\ friend, face, talking, bitch, court \end{tabular}  &  0.055 \\ \midrule
Topic 07       & \begin{tabular}[c]{@{}l@{}}sport, basketball, school, city, jersey,\\ high, black, world, college, white \end{tabular}  &  0.054\\ \midrule
Topic 08       & \begin{tabular}[c]{@{}l@{}}coach, bench, starting, role, system,\\ fit, coaching, front, starter, minute \end{tabular}  &  0.054 \\ \midrule
Topic 09    & \begin{tabular}[c]{@{}l@{}}prime, greatest, goat, time, career, \\ all, star, history, seasons, era \end{tabular}             &   0.043  \\ \bottomrule
\end{tabular}
\caption{The remaining 10 extracted topics by LDA using all comments in \communityname{/r/NBA}. The top ten weighted words are presented for each topic.}
\label{tab:tentopic}
\end{table}

\subsection{Game Thread Matching}
\label{sec:appendix_thread}

The percentage of game threads detected among all games by season is shown in Table~\ref{tab:detect}. 
The percentages of game threads detected in \communityname{/r/NBA} are high in all seasons, 
with an average of 95\%. 
The percentage of game threads detected in team subreddits are high in the 2016 and 2017 seasons, all above 80\%. 
Significant amount of game threads are missing for the 2013 and 2014 seasons. 

We further investigate the reasons why game threads are not detected in team subreddits. 
We randomly sampled 50 games in season 2014. 
If home-team game thread and away-team game thread exist for these games,
100 game threads should have been detected in total. 
We manually checked them in our dataset and found that 40 game threads are successfully detected, 
54 game threads were not created and 6 game threads had special title formats. 
Based on this limited sample, our detection rate is in fact 87\% (40/46).
There are two major reasons to explain missing game threads:
1) Some team subreddits were relatively small and fans only created game threads for important games 
(e.g., games against rivalry teams and strong teams, or games that are critical for playoff spots); 
2) Some Game Threads do not follow the standard format. For example, 
a game thread in \communityname{/r/Timberwolves} is 
titled ``Last regular season game, boys travel to Houston!''.

\begin{table}[]
\centering
\small
\begin{tabular}{l|rrrrr|rrrrr}
\toprule
                          & \multicolumn{5}{c|}{Team subreddits} & \multicolumn{5}{c}{\communityname{/r/NBA}}          \\
\multicolumn{1}{c|}{year} & 2013  & 2014  & 2015  & 2016 & 2017  & 2013 & 2014 & 2015  & 2016  & 2017  \\ \midrule
\#game threads detected    & 503   & 1176   & 2131  & 2362 & 2424  & 1106 & 1230 & 1311  & 1314  & 1305  \\
\#games                   & 1314  & 1319  & 1311  & 1316 & 1309  & 1314 & 1319 & 1311  & 1316  & 1309  \\
Percentage                & 19\%  & 45\%  & 81\%  & 90\% & 93\% & 84\% & 93\% & 100\% & 100\% & 100\% \\ \bottomrule
\end{tabular}
\caption{Percentage of game threads detected from team subreddits and \communityname{/r/NBA} by season. 
For each game, it is supposed to have one game thread in home-team subreddit, 
one game thread in away-team subreddit and one game thread in \communityname{/r/NBA}. 
The detected game thread is high for \communityname{/r/NBA}. 
For team subreddits, certain percentage of Game Thread are missing because: 
1) Team subreddits only create game threads for important games when the community is relatively small; 
2) Some game threads do not follow standard format.}
\label{tab:detect}
\end{table}

\subsection{\added{F-tests with Bonferroni correction}}
\label{sec:ftest}
\added{Table~\ref{tab:ftest} summarizes the results of F-tests where the null hypothesis is that adding team performance variables does not provide a significantly better fit. All the $p$-values are less than 0.0001 after the 
Bonferroni correction, so we reject the null hypothesis.
In all the regressions in Table~\ref{tab:ftest}, adding team performance variables provides a significantly better fit.}
\begin{table}[]
\small
\begin{tabular}{l|cc|cc|cc}
\toprule
        & \multicolumn{2}{c|}{H1}   & \multicolumn{2}{c|}{H2}        & \multicolumn{2}{c}{H3}                 \\ 
        & \multicolumn{1}{c}{Team subreddits}   & \multicolumn{1}{c|}{\communityname{/r/NBA}}  & \multicolumn{1}{c}{Seasonly} & \multicolumn{1}{c|}{Monthly} & \multicolumn{1}{c}{\topicname{season prospects}} & \multicolumn{1}{c}{\topicname{future}}          \\ \midrule
$F$-value & 26.9              & 51.6              & 15.2                    & 12.3                   & 16.3               & 20.1              \\ 
$p$-value & \textless{}0.0001 & \textless{}0.0001 & \textless{}0.0001       & \textless{}0.0001      & \textless{}0.0001  & \textless{}0.0001 \\ \bottomrule
\end{tabular}
\caption{\added{$F$-test results.
All the $p$-values are less than 0.0001 after the Bonferroni correction.}}
\label{tab:ftest}
\end{table}

\subsection{Additional Linear Regression Models for Topic Weights}
\label{sec:appendix_topic_regression}

Table~\ref{tab:3topics} presents the results of OLS linear regression models for average topic weights of top five topics other than \topicname{future} and \topicname{season prospects,} i.e., \topicname{personal opinion,} \topicname{game strategy,} and \topicname{game stats.}
\begin{table}[t]
\small
\centering
\begin{tabular}{l|LL|LL|LL}
\toprule
\multicolumn{1}{l|}{}   & \multicolumn{2}{c|}{\topicname{personal opinion}}                    
& \multicolumn{2}{c|}{\topicname{game strategy}}  & \multicolumn{2}{c}{\topicname{game stats}}                                                                                                 \\
\multicolumn{1}{c|}{variables}              & \multicolumn{1}{c}{Reg. 1}                   & \multicolumn{1}{c|}{Reg. 2}          & \multicolumn{1}{c}{Reg. 1}                     & \multicolumn{1}{c|}{Reg. 2}                    & \multicolumn{1}{c}{Reg. 1}    
& \multicolumn{1}{c}{Reg. 2}                            \\ \midrule
\textit{Control: season}  &&&&&&\\
2014        			& 0.110*** 	& 0.125***  	& 0.079*** 	& 0.082***	& 0.059***	& 0.059***	\\
2015   					& 0.146***  & 0.196***  	& 0.096*** 	& 0.095***	& 0.083***	& 0.082***	\\
2016       				& 0.063**  	& 0.107*** 		& 0.101***	& 0.090***	& 0.050***	& 0.088***	\\
2017      				& 0.152***  & 0.211***  	& 0.089*** 	& 0.081***	& 0.104***	& 0.103***	\\[5pt]
\textit{Performance}  	&&&&&&\\
season elo         		& 			& -0.380**		& 		 	& -0.126	& 			& -0.099 \\
season elo difference   & 			& -0.125*  		& 		 	& -0.024	&			& 0.044	\\[5pt]
\textit{Team information} &&&&&&\\
\multicolumn{1}{l|}{market value}          &   & 0.080   & 	  & 0.015  & 		& 0.014 \\
\multicolumn{1}{l|}{average age}          &   & -0.121*   &   & -0.069*  & 		& -0.053* \\
\multicolumn{1}{l|}{\added{\#star players}} &   & 0.018   &   & -0.068  & 	& -0.043  \\
\multicolumn{1}{l|}{\added{\#unique users}} &   & -0.105   &   & 0.068  & 		& 0.146*** \\
\multicolumn{1}{l|}{offense points}        &  & 0.158   &   & 0.112  & 		& -0.100 \\
\multicolumn{1}{l|}{defense points}        &   & -0.213*   &   & 0.038  & 		& 0.053 \\
\multicolumn{1}{l|}{turnovers}        &   & -0.120   &   & 0.032  & 		& -0.079 \\
\midrule
\multicolumn{1}{l|}{intercept}       &   & 0.781***   &   & 0.414***  & 0.043 & 0.414*** \\ 
\multicolumn{1}{l|}{Adjusted $R^2$} & 0.048	& 0.208	& 0.008	& 0.251	& 0.025	& 0.334	\\
\bottomrule
\end{tabular}
\caption{\replaced{Hierarchical regression analyses}{Linear regression models} for the other top 5 topics: \topicname{personal opinion,} \topicname{game strategy,} and \topicname{game stats.} }
\label{tab:3topics}
\end{table}

\end{document}